\documentclass[aps,prd,groupedaddress,nofootinbib,preprintnumbers]{revtex4}
\usepackage{graphicx}
\usepackage{amsfonts}
\usepackage{amssymb}
\usepackage{amsmath}
\usepackage{tabularx,url,color}
\usepackage{hyperref}
\usepackage{multirow}

\newcommand{\Mzero}{\ensuremath{m_0}}
\newcommand{\MOneHalf}{\ensuremath{m_{1/2}}}
\newcommand{\Azero}{\ensuremath{A_0}}
\newcommand{\tanb}{\ensuremath{\tan\beta}}
\newcommand{\omegacdm}{\ensuremath{\Omega_\text{cdm}h^2}}
\newcommand{\Mtop}{\ensuremath{m_t}}
\newcommand{\Mone}{\ensuremath{M_1}}
\newcommand{\Mtwo}{\ensuremath{M_2}}
\newcommand{\Mthree}{\ensuremath{M_3}}
\newcommand{\Deltaamu}{\ensuremath{\mathrm{\Delta a_\mu}}}

\newcommand{\chisq}{{\ensuremath -2\log L}}

\def\eg{\textit{e.g.}\ }
\def\ie{\textit{i.e.}\ }
\def\etal{\textit{et al}\ }

\begin{document}

\title{Constraining Supersymmetry using the relic density and the Higgs boson}

\author{Sophie Henrot-Versill\'e$^1$, 
        R\'emi Lafaye$^2$, 
        Tilman Plehn$^3$, 
        Michael Rauch$^4$,  
        Dirk Zerwas$^1$,
	St\'ephane Plaszczynski$^1$, 
	Benjamin Rouill\'e d'Orfeuil$^1$, 
        Marta Spinelli$^1$}

\affiliation{$^1$LAL, CNRS/IN2P3, Orsay Cedex, France}
\affiliation{$^2$LAPP, Universit\'e de Savoie, IN2P3/CNRS, Annecy, France}
\affiliation{$^3$Institut f\"ur Theoretische Physik, Universit\"at Heidelberg, Germany}
\affiliation{$^4$Institute for Theoretical Physics, Karlsruhe Institute of Technology (KIT), Karlsruhe, Germany}

\begin{abstract}
Recent measurements by Planck, LHC experiments, and Xenon100 have 
significant
impact on supersymmetric models and their parameters.  
We first illustrate the constraints in the mSUGRA plane
and then perform  a detailed analysis of the general MSSM with 13 free parameters.
Using SFitter, Bayesian and Profile Likelihood approaches are applied and their results compared.
The allowed structures
in the parameter spaces are largely defined by different mechanisms of
dark matter annihilation in combination with the light Higgs mass
prediction. In mSUGRA the pseudoscalar Higgs funnel and stau
co-annihilation processes are still avoiding experimental pressure. In
the MSSM stau co-annihilation, the light Higgs funnel, a mixed
bino--higgsino region including the heavy Higgs funnel, and a large
higgsino region predict the correct relic density.  Volume effects and
changes in the model parameters impact the extracted mSUGRA and MSSM
parameter regions in the Bayesian analysis.
\end{abstract}

\maketitle

\tableofcontents
\newpage

%%%%%%%%%%%%%%%%%%%%%%%%%%%%%%%%%%%%%%%%%%%%%%%%%%%%%%%%%%%%%%%%%%%%%%%%%%%%%%%%
\section{Introduction}

When trying to understand the physics of the electroweak scale we
encounter a set of experimental and theoretical problems. First, the
discovery of a narrow, most likely fundamental Higgs scalar means that
the hierarchy problem is now real~\cite{higgs}. Second, dark matter
search experiments like Xenon100 are starting to cut into the available
parameter space of a weakly interacting dark matter
particle~\cite{xenon}. Third, in the 7~TeV and 8~TeV runs there seems
to be no hint for any physics beyond the Standard Model at ATLAS, CMS,
and LHCb.

On the other hand, the particle nature of dark matter is
the most attractive hypothesis. The key observable in such models is the current dark
matter density in the Universe. The Planck collaboration has recently
released their data on the cosmological microwave background
temperature anisotropies~\cite{Planckdata}. In the $\Lambda$CDM
scenario they determine the dark matter density $\omegacdm$ with an
unprecedented accuracy~\cite{Planck:param}.\bigskip

If we take for example the approximate gauge coupling
unification~\cite{coupling_uni} as a motivation to look for a weakly
interacting ultraviolet completion of the Standard Model we are still
driven to supersymmetry~\cite{susy}. The discovery of a Higgs boson in
the Minimal Supersymmetric Standard Model (MSSM) mass range~\cite{m_h} gives no reason to modify or abandon
this hypothesis. Moreover, alternative structures like extra
dimensions or little--Higgs--like models find themselves under at
least as much experimental pressure as supersymmetry. The question we
should answer at the end of the first phase of LHC is which part of
the MSSM parameter space is consistent with all available
(non)-observations.\bigskip

Even in supersymmetric models the nature of dark matter remains an
open question. While in the MSSM the only available TeV-scale dark
matter candidate is the lightest neutralino~\cite{neutralino}, very
weakly interacting dark matter particles might exist at much lower
masses~\cite{axino}. In the light of many indirect constraints on
neutralino--induced higher--dimensional operators we can extend the
dark matter fermion to a Dirac spinor~\cite{rsymmetry}, predicting
interesting but still unobserved sgluon signatures at the
LHC~\cite{sgluon}. Recent years have seen a large effort to condense
properties of different dark matter models into effective theory
concepts~\cite{tim}. In spite of all these options we will limit
ourselves to the case where the entire observed dark matter density is
due to a single state, the lightest Majorana neutralino. If this
hypothesis comes under experimental pressure, this might serve as
a motivation for more elaborate dark sectors, but we will see that
there is no such pressure.\bigskip

Our analysis is based on the \textsc{SFitter} toolkit which determines
the underlying parameters of complex models in the absence of simple
one--to--one correlations of observables and parameters. We explore 
the parameter space with Monte Carlo Markov Chains of
the likelihood function. 
This tool also permits to compare the results within the same framework, 
using either Bayesian or Profile Likelihoods analysis.
It has
previously been applied to the problem of the determination of
supersymmetric parameters~\cite{sfitter_release}, including a
bottom-up renormalization group analysis and experimental information
on production rates~\cite{sfitter_susy}, as well as Higgs coupling
measurements~\cite{sfitter_higgs}. In this paper we will study the
impact of the recent LHC Higgs measurements and of the $\omegacdm$
measurement by Planck. We will compare the latter to the WMAP-9year
results~\cite{Wmap:param}. 

We will use mSUGRA~\cite{msugra} as
an illustration of the constraints than can be put through the use
of this full set of measurements. This step is necessary to study what
happens in models where the a-priori relatively unrelated weak dark
matter sector, Higgs sector, and strongly interacting sector of the
MSSM are strongly linked by a high-scale construction.
The main emphasis of our study is the challenging study of a TeV-scale MSSM. The determination
of its parameters is the ultimate goal in order to infer from data 
whether the parameters are unified at a higher scale. This determination should
shed light on which scenarios of SUSY breaking might be favored~\cite{sfitter_susy}.
Therefore we use a 13 parameter MSSM, which is a technically challenging endeavor because
of the large number of parameters. In addition we use the top mass as an input
and as a parameter. This additional parameter helps fine tuning
the Higgs mass for dark matter annihilation.

Similar studies have been performed by other groups considering different
models: for instance,
\textsc{Fittino} has studied the impact of LHC data and WMAP-7year
results~\cite{fittino} on two models, mSUGRA and a non-universal Higgs
model, the \textsc{MasterCode} group has performed a likelihood study of
the same mSUGRA and non-universal Higgs
models including Xenon100 results~\cite{Buchmueller:2012hv}. A
specific analysis with Planck data, the Higgs mass measurement, and
Xenon100 in the TeV-scale MSSM exists but which focuses on light neutralino
dark matter~\cite{boehm}. Results similar to ours have recently been
published in Ref.~\cite{jay} for mSUGRA and by the BayesFITS group, including 
the study of a 9-parameter MSSM in Ref~\cite{BayesFITS}. 
Compared to this model, we are letting the data constrain more 
parameters, rendering the determination more complex.
A non-exhaustive list of other, similar
analyses is given in Ref.~\cite{DMpapers}.

%%%%%%%%%%%%%%%%%%%%%%%%%%%%%%%%%%%%%%%%%%%%%%%%%%%%%%%%%%%%%%%%%%%%%%%%%%%%%%%%
\section{Supersymmetric parameters}
\label{sec:models}

With the limited number of actual measurements entering this analysis
it is clear that we will not be able to make any definite statements
about a full TeV-scale supersymmetric mass spectrum. We will
illustrate our results using two model setups. As a first test we will
study the unified gravity--mediated mSUGRA model, this
will give us some ideas about how strongly unified models can
accommodate the various data constraints. Second, in a proper
bottom--up approach we will consider a free TeV-scale spectrum, reduced
to the subset of relevant mass parameters.\bigskip

The strongly constrained mSUGRA model is described by three mass
parameters defined at the unification scale: $\Mzero$, the common
scalar breaking mass parameter, $\MOneHalf$ the common gaugino
breaking mass parameter and $\Azero$, the common trilinear mass
parameter. In addition, $\tanb$ as the ratio of the vacuum expectation
values of the two Higgs doublets encodes successfully electroweak
symmetry breaking.  Finally, we have to fix the sign of the higgsino
mass parameter $\mu$. In our conventions the term $-\mu$ appears in
the lower-right off-diagonal terms of the neutralino mass matrix. The
off-diagonal entry in the stop mass matrix is $m_t (A_t - \mu \cot
\beta)$~\cite{suspect}. Because the parameter $A_t$ is the key
parameter in the computation of the light Higgs mass around 126~GeV we
quote the approximate solution to the renormalization group
evolution~\cite{drees_martin},
\begin{alignat}{5}
A_t 
&= A_0 \left( 1 - \frac{0.75}{\sin^2 \beta} \right) \; 
 - 3.5 \MOneHalf \left( 1 - \frac{0.41}{\sin^2 \beta} \right) 
&\approx \begin{cases}
   0.62 A_0 - 2.8 \MOneHalf 
 & \text{for} \tan \beta = 1 \\[2mm]
   0.25 A_0 - 2.1 \MOneHalf 
 \qquad & \text{for} \tan \beta \gg 1 \; .
   \end{cases}
\label{eq:msugra_at}
\end{alignat}
The larger $\tan \beta$ becomes the more the weak-scale parameter
$A_t$ is driven by $\MOneHalf$. For $\MOneHalf > 0$ we essentially
almost find $A_t < 0$.\bigskip

When we use renormalization group equations to run high--scale
supersymmetry breaking parameters to the weak scale, fixed to 1~TeV
as suggested by Ref.~\cite{EPJC}, we need to ensure
that we successfully generate the observed electroweak symmetry
breaking. It is convenient to include $\tan \beta$ as a mSUGRA model
parameter, but this choice mixes high--scale mass parameters with a
TeV--scale ratio of vacuum expectation values.  To be more consistent
in the definition of the mSUGRA parameter space we can avoid $\tan
\beta$ and replace it with the appropriate mass parameters evaluated
at the unification scale~\cite{drees_martin},
\begin{alignat}{5}
\mu^2 &=  \frac{m^2_{H_u} \sin^2 \beta - m^2_{H_d} \cos^2 \beta}
               {\cos (2 \beta)}
        - \frac{1}{2} m_Z^2 \notag \\
2 B \mu &= \left( m^2_{H_d} - m^2_{H_u} \right) \, \tan (2 \beta) 
           + m_Z^2 \, \sin (2 \beta) \; .
\end{alignat}
$H_u$ has a tree--level coupling to
up--type fermions, while $H_d$ couples to down--type fermions. The
parameter $B \mu$ accompanies the doublet mixing $H_u^0
H_d^0$. Instead of $m_{H_j}$ and $\tan\beta$ we can use $B$ and $\mu$
and the correct value of $m_Z$ as mSUGRA model parameters. For the
profile likelihood approach the two parametrizations are equivalent. However,
for the Bayesian approach they will lead to different priors and hence
to different results~\cite{sfitter_release}. In terms of $\tan \beta$ a flat
prior in the high--scale mass parameters corresponds to the
prior~\cite{Allanach:2007qk}
\begin{equation}
\left|
\frac{m_Z}{2\mu^2} \;
\left(m_{H_u}^2+m_{H_d}^2+2\mu^2 \right) 
\frac{1-\tan^2 \beta}{(1+\tan^2 \beta)^2} \;
\right| \; ,
\label{eq:jacobian}
\end{equation} 
defined at the electroweak scale. At large values of $\tan \beta$, the
Jacobian in Eq.\eqref{eq:jacobian} scales like $1/\tan^2 \beta$, which
means that the high--scale flat prior prefers small values of
$\tan \beta$. When we define the entire MSSM parameter set
at the TeV--scale, Jacobians like the one shown in
Eq.\eqref{eq:jacobian} are simply an effect of our freedom to choose
our MSSM model parameters.\bigskip

In computing the weak--scale mass spectrum in the simple mSUGRA model
we start with the independent GUT--scale parameters, evolve the soft
SUSY--breaking parameters to the TeV scale, and compute the
corresponding masses of the supersymmetric states. \textsc{SFitter}
primarily relies on \textsc{SuSpect2}~\cite{suspect} for the
renormalization group evolution and the computation of the
supersymmetric mass spectrum. In addition, we use
\text{SoftSUSY}~\cite{softsusy} to test our results. Because of the
different behavior of the squark and the gaugino masses in the
$\Mzero$ vs $\MOneHalf$ plane~\cite{Jaeckel:2011wp} some complexity of
the mSUGRA model arises through parameter correlations.\bigskip

The most general MSSM contains a large number of parameters, of which
we identify 17 which will affect current LHC and dark matter
measurements~\cite{sfitter_release}. Moreover, the absence of evidence
for supersymmetric particles at the LHC leads us to effectively
decouple some of the masses to values well about the TeV scale.

In this analysis all squark mass parameters with the exception of the
stop sector are fixed at 2~TeV. The same value is assumed for the
gluino mass parameter $\Mthree$.  This way gluinos and light--flavor
squarks move outside the region excluded by the LHC. The question of
the bias introduced by this assumption will be addressed later.  The
trilinear mass parameter $A_b$ is assumed to be zero. The
first--generation slepton parameters are identified with their
second--generation counter parts. This leaves 13~supersymmetric
parameters to be explored: $\tan\beta$, the electroweak gaugino mass
parameters ($\Mone$, $\Mtwo$), the smuon and stau sectors
($M_{\tilde{\mu}_{L,R}}$, $M_{\tilde{\tau}_{L,R}}$, $A_\tau$), the
stop sector ($M_{\tilde{q}_{3L}}$, $M_{\tilde{t}_R}$, $A_t$), the heavy Higgs mass
$m_A$, and the higgsino mass parameter $\mu$.

Effectively, this reduced parameter space decouples the strongly
interacting MSSM sector from the weak sector with the relevant dark
matter and Higgs predictions. The only remaining strongly interacting
particle in the picture is the top squark with its large impact on the
Higgs sector --- related to its particular relevance in the solution
of the hierarchy problem. Since the uncertainties in the top quark
mass are non-negligible, and because the induced parametric
uncertainties for example for the light MSSM Higgs mass cannot be
neglected, we include it as an additional model parameter in the
mSUGRA as well as in the MSSM analysis.

The prediction of
the light MSSM Higgs mass is calculated with
\textsc{SuSpect2}~\cite{suspect} while the Higgs branching ratios are
computed using \textsc{Susy-Hit} and \textsc{HDecay}~\cite{s-hit}. The
supersymmetric contribution to the cold dark matter density is
calculated with \textsc{MicroMegas}~\cite{Belanger:2010gh}. For the
electroweak precision observables we rely on
\textsc{SusyPope}~\cite{Heinemeyer:2007bw}. Finally, we use
\textsc{SuSpect2}~\cite{suspect} and
\textsc{MicroMegas}~\cite{Belanger:2010gh} to compute the $B$
observables and $(g-2)_\mu $.

%%%%%%%%%%%%%%%%%%%%%%%%%%%%%%%%%%%%%%%%%%%%%%%%%%%%%%%%%%%%%%%%%%%%%%%%%%%%%%%%
\section{Annihilation channels}
\label{sec:ann}

To study the effect of the measured dark matter relic density on the
supersymmetric parameter space we need to take into account the fact that 
data
drive us into a few distinct parameter regimes~\cite{jay}. These
structures combined with the light Higgs mass prediction will lead to
well--defined regions of the mSUGRA and MSSM parameter spaces which
are consistent with all current data.\bigskip

The first is the light Higgs funnel region where the mass of the
lightest Higgs boson is about twice the mass of the LSP. The leading
contribution to dark matter annihilation is then the $s$-channel
annihilation via the lightest Higgs, dominantly decaying to $b$
quarks. As a consequence of the tiny width of the lightest Higgs, $\Gamma_h
\sim 5$~MeV, the LSP mass has to be finely adjusted to produce the
correct range in $\omegacdm$. A small, $\mathcal{O}(10\%)$ higgsino
component of the LSP will give the correct relic density. Technically,
this precise tuning will be a challenge for our parameter analysis.

The same $s$-channel annihilation can proceed via the heavy Higgs
bosons $A,H$, where the widths can be very large and the level of
tuning will be smaller. Unlike the $h$-funnel, this $A$-funnel region
can extend to arbitrarily large LSP masses, provided the Higgs masses
follow the LSP mass. The main heavy Higgs decay channels are
$b\bar{b}$ and $t\bar{t}$, because, in these kinds of two-Higgs-doublet
models, the massive gauge bosons decouple from the heavy Higgs sector.

A second annihilation topology gives rise to the $\tau$
co-annihilation region~\cite{stau-co-annihilation}. Here, the mass
difference between for example the stau as the next-to-lightest
supersymmetric particle and the LSP needs to be small, of the order of
few per-cent or less.  If the LSP has a large higgsino component the
annihilation then proceeds via an $s$-channel tau lepton into a tau
and a pseudo--scalar Higgs. On the scale of the size of LHC detectors
the stau could in such scenarios become stable. However, the higgsino
component is not required for co-annihilation to work.  If instead the
selectron or smuon are the next-to-lightest supersymmetric particles
and essentially mass degenerate with the LSP they could lead to the same
effect. In the squark sector the same mechanism exists for the
lightest top squark~\cite{stop-co-annihilation} or other squark
next-to-lightest superpartners. However, given the preference of the
Higgs mass measurement for heavy stop masses we find it outside our
preferred parameter range.

In the absence of a significant mass splitting between the lightest
neutralino and lightest chargino, co-annihilation in the
neutralino--chargino sector can accelerate dark matter annihilation in
the early universe~\cite{char-co-annihilation}. Because the necessary
mass degeneracy cannot appear for a light bino, the LSP will be
dominantly wino or higgsino. Two final states occur for
neutralino--chargino co-annihilation: if the $t$-channel neutralino or
chargino exchange dominates, massive gauge bosons and eventually
light--flavor quarks will be produced in the annihilation process. If,
in contrast, a heavy Higgs in the $s$-channel dominates, the final
state will dominantly consist of third-generation quarks.

Finally, the focus point region~\cite{drees_martin,focus_point} is
characterized by large $\Mzero$, small $\MOneHalf$, and accidentally
small $|\mu|$. Close to this region of parameter space where $\mu$
changes sign, we find a higgsino-like light neutralino which couples
to gauge bosons and can annihilate into the $WW$ channel. For mSUGRA,
this area is highly reduced by both Xenon100~\cite{Buchmueller:2012hv}
and LHC gluino search limits~\cite{Aad:2012fqa,hCMS}.
 
%%%%%%%%%%%%%%%%%%%%%%%%%%%%%%%%%%%%%%%%%%%%%%%%%%%%%%%%%%%%%%%%%%%%%%%%%%%%%%%%
\section{Data and analysis setup}
\label{sec:setup}

%--------------------------------------------------
\begin{table}[t]
  \centering
  \begin{tabular}{l|c c}
\hline
measurement & value and error & \\
\hline
$m_h$                    & $(126\pm 0.4 \pm 0.4 \pm 3)$~GeV           & \cite{Higgsmass} \\
\hline
$\omegacdm$ Planck       & $0.1187\pm 0.0017 \pm 0.012$      & \cite{Planck:param} \\
$\omegacdm$ WMAP-9year &  $0.1157 \pm 0.0023 \pm 0.012$    & \cite{Wmap:param} \\
\hline
BR($B_s \to \mu^+\mu^-$) & $(3.2 ^{+1.5}_{-1.2} \pm 0.2)\times 10^{-9}$ & \cite{Bsmumu} \\ 
%BR($B_s \to \mu^+\mu^-$) & $(2.9 ^{+1.1}_{-1.0} \pm 0.2)\times 10^{-9}$ & \cite{Bsmumu2} \\ 
BR($b \to X_s \gamma$)  & $(3.55 \pm 0.24 \pm 0.09) \times 10^{-4}$   & \cite{XsGamma} \\
$\Deltaamu$             & $(287 \pm 63 \pm 49 \pm 20) \times 10^{-11}$& \cite{Amu} \\ 
$\Mtop$                 & $(173.5\pm 0.6\pm 0.8)$~GeV                   & \cite{Topmass} \\
\hline
  \end{tabular}
\caption{Some of the key measurements used in our analysis, including
  the error. The last number is the theoretical uncertainty on the
  supersymmetric prediction, except for the BR($b \to X_s \gamma$) and
  $m_t$ for which no theoretical uncertainty is considered. }
\label{tab:data}
\end{table}
%--------------------------------------------------

In Table~\ref{tab:data} we list the main experimental inputs to our
analysis.  The Higgs mass measurement at the LHC considered in this
study is from ATLAS~\cite{Higgsmass}. Because it comes with a sizeable
theoretical error from the supersymmetric prediction an improved
measurement, such as the measurement of $BR(B_s \to \mu^+\mu^-)$ 
by CMS~\cite{CMSBsmumu}, will not affect our results. The different production and
decay channels of the Higgs boson~\cite{ATLAS:2012ac} provide some
additional information on its couplings~\cite{sfitter_higgs}, but with
little impact on the supersymmetric parameter space when added to the
Higgs mass and the flavor observables~\cite{sven_etal}. We consider
the updated result $\text{BR}(B_s \to \mu^+\mu^-)=(2.9 ^{+1.1}_{-1.0}
\pm 0.2)\times 10^{-9}$~\cite{Bsmumu2} in the mSUGRA section. It has
no impact on the results so we stick to the value quoted in
Table~\ref{tab:data} for the MSSM study.\bigskip

A major point of this study is on the new measurement of the cold dark
matter density of the universe by the Planck collaboration. We compare
it with the WMAP-9year measurement.
%Several $\omegacdm$ measurements are provided by both
%collaborations combining their own CMB dataset with others.  
In both cases we use the values from the more precise measurements in
the $\Lambda$CDM scenario:
\begin{itemize}
\item[--] Planck: \quad $\omegacdm=0.1187\pm 0.0017$~\cite{Planck:param}
  \\ This is a combination of Planck data, large scale polarization
  WMAP data~\cite{Wmap-polar}, ACT/SPT~\cite{ACTSPT}, and baryon
  acoustic oscillation measurements (BAO)~\cite{BAO}.
\item[--] WMAP-9year: \quad $\omegacdm=0.1157 \pm
  0.0023$~\cite{Wmap:param} \\ This combines WMAP data,
  BAO and a Hubble parameter measurement~\cite{H0}.
\end{itemize}
Some tension remains between Planck's estimated $H_0$ value and the
direct measurements used in the WMAP-9year analysis. We compare 
the two approaches to see whether the difference in central values
and errors leads to differences in the constraints on the
supersymmetric parameter space.

An additional dark matter related input is the upper limit on the
elastic LSP--Nucleon cross section as function of the LSP mass from
the analysis of the Xenon100 225~days $\times$~34~kg
dataset~\cite{xenon}.\bigskip

For $\tan\beta>50$ the branching ratio of the flavor violating decay
$B_s \to \mu^+\mu^-$ is particularly sensitive to supersymmetric
contributions~\cite{bsmumuMSSM} and hence constraining.  The
measurement of BR($b\to X_s \gamma$) tends to disfavor $\mu<0$ for
large $\tan\beta$~\cite{BXsMSSM}. The difference in the predicted and
measured anomalous magnetic moment of the muon tends to accommodate
large $\tan\beta$ and to disfavor
$\mu<0$~\cite{muMSSM}. The reason for this definite sign preference
in $\mu$ is a possible cancellation in the off-diagonal entries of the
third generation scalar mass matrices. The top mass~\cite{Topmass} is
both, a model parameter and a measurement.\bigskip

In \textsc{SFitter} the statistical errors on the measurements are
treated as Gaussian or Poisson where appropriate. The systematic
errors are correlated if originating from the same source. Theoretical
uncertainties are treated with the \textsc{Rfit}
scheme~\cite{rfit,sfitter_release,sfitter_higgs}, \ie using
flat errors in a profile likelihood construction.

The analysis proceeds in two steps. First, we construct a
fully exclusive log-likelihood map in the model parameters using a set of
Markov Chains with a Breit-Wigner proposal function.  Each chain has a
different starting point.  Their convergence is checked by comparing the
mean values and variances of each chain through the quantity
$\hat{R}$~\cite{Gelman:1992zz} as implemented in
Ref.~\cite{Allanach:2005kz}.  The maximum over the set of Markov
Chains $\max[\hat{R}]$ will approach unity if the chains have converged
and cover the full parameter space.

On this exclusive log-likelihood map we then define two types of
projections: a profile likelihood based on the Frequentist approach
and a marginalization as an example of the Bayesian approach.  The
absolute scales of the projected log-likelihood values in the two
approaches can not be used to compare them.

In 2-dimensional standard contour plots we identify the interesting parameter
regions and their correlations.  In these regions we explore the
structures locally, using a modified version of
\textsc{Minuit}~\cite{minuit} to refine the location of the minima. 

%%%%%%%%%%%%%%%%%%%%%%%%%%%%%%%%%%%%%%%%%%%%%%%%%%%%%%%%%%%%%%%%%%%%%%%%%%%%%%%%
\section{mSUGRA analysis}
\label{sec:msugra}

The strongly constrained mSUGRA parameter space is governed by a very
small number of parameters. They are linked to the TeV-scale masses
via coupled renormalization group running and therefore highly
correlated. Knowing the light Higgs mass further constrains the
parameter space through the stop sector.  In such cases the standard
Markov Chain Monte Carlo methods give the most stable results, so we
do not use the weighted Markov Chains which are otherwise optimized
for a small number of parameters~\cite{sfitter_release,wmc}.  For
each sign of $\mu$ we travel in the 5-dimensional parameter space of
$\Mzero$, $\MOneHalf$, $\Azero$, $\tanb$, and $m_t$ with 49 Markov
chains of 200000 points each, giving us 9.8~million accepted
samplings.  Our parameter space is bounded by $\Mzero<5$~TeV,
$\MOneHalf<5$~TeV, $|\Azero|<4$~TeV, and $\tanb<61$. The convergence
criterion finds $\max[\hat{R}] \approx 1.008$, indicating a good
convergence of the chains.

%%%%%%%%%%%%%%%%%%%%%%%%%%%%%%%%%%%%%%%%%%%%%%%%%%%%%%%%%%%%%%%%%%%%%%%%%%%%%%%%
\subsection{Profile likelihood for positive $\mu$}

%--------------------------------------------------
\begin{figure}[t]
\includegraphics[width=0.49\textwidth]{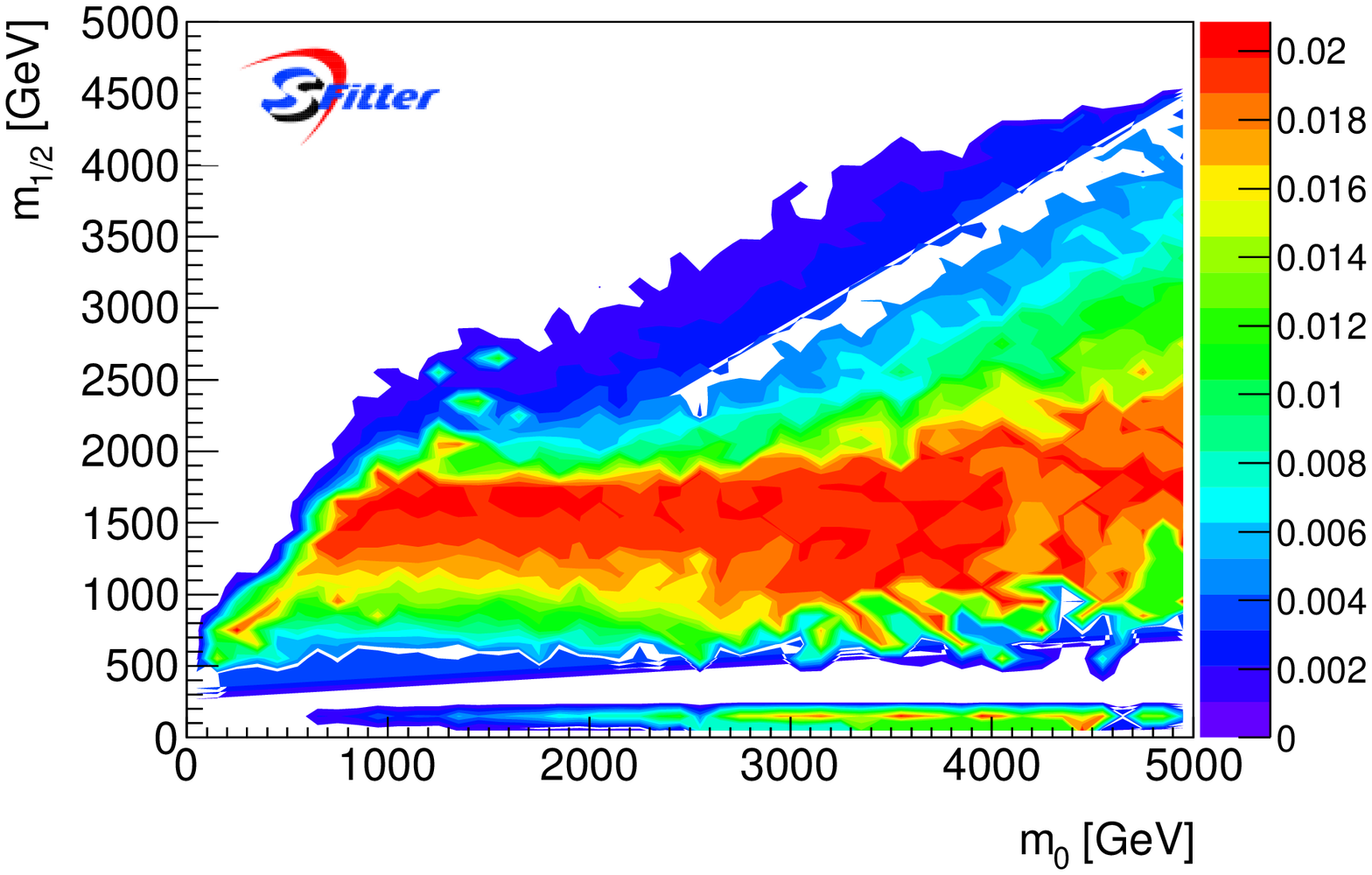}
\includegraphics[width=0.49\textwidth]{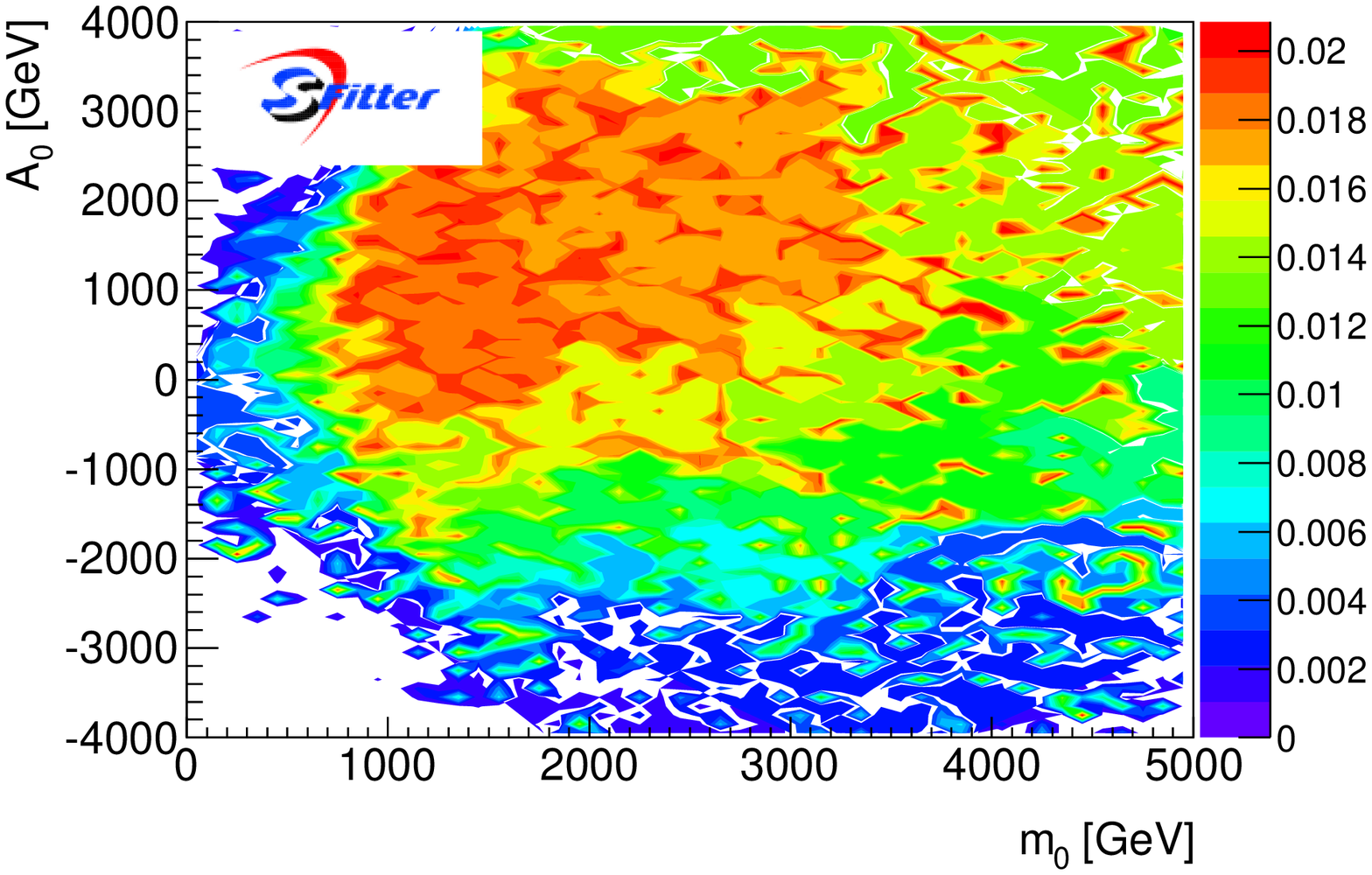} \\
\includegraphics[width=0.49\textwidth]{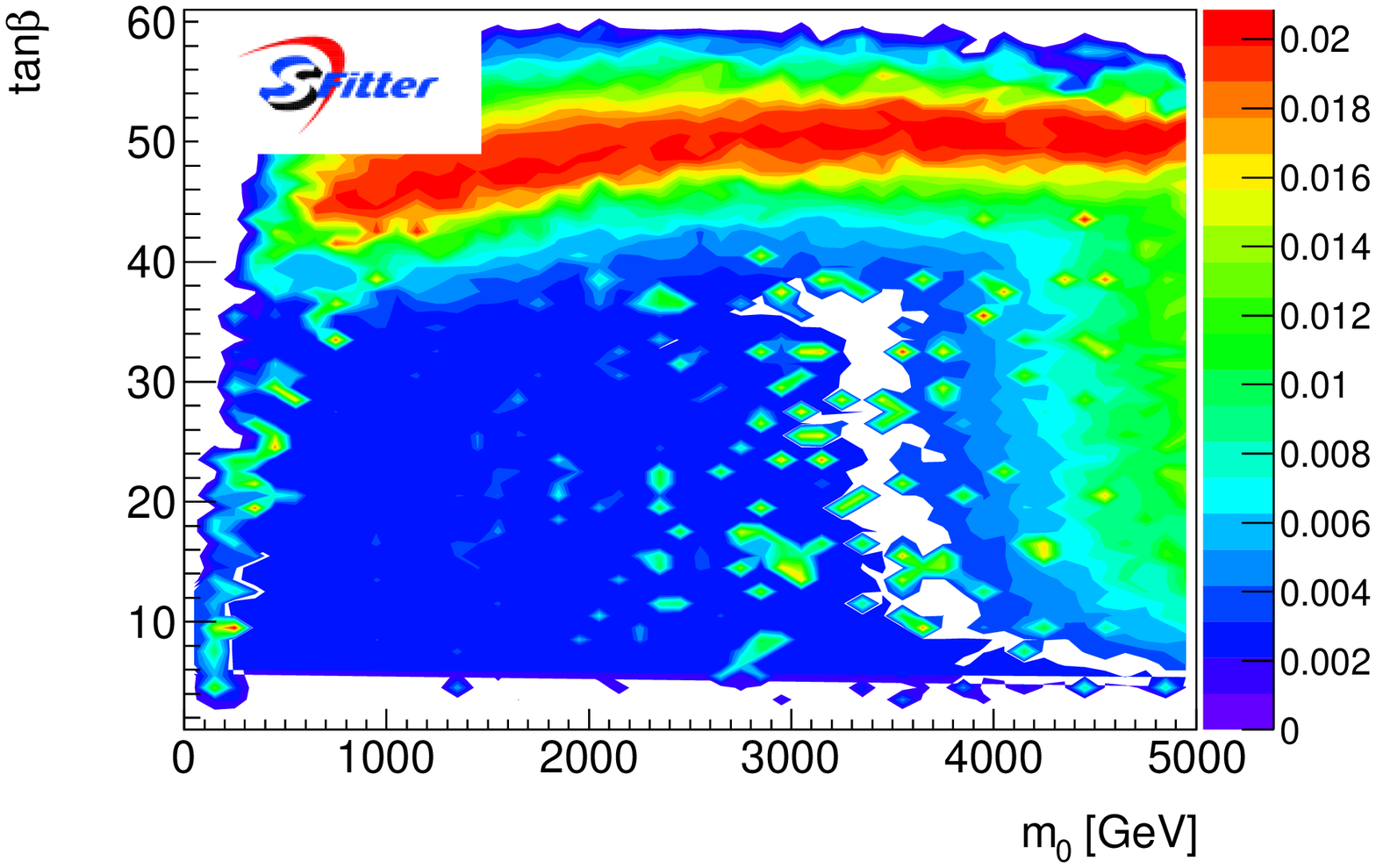}
\includegraphics[width=0.49\textwidth]{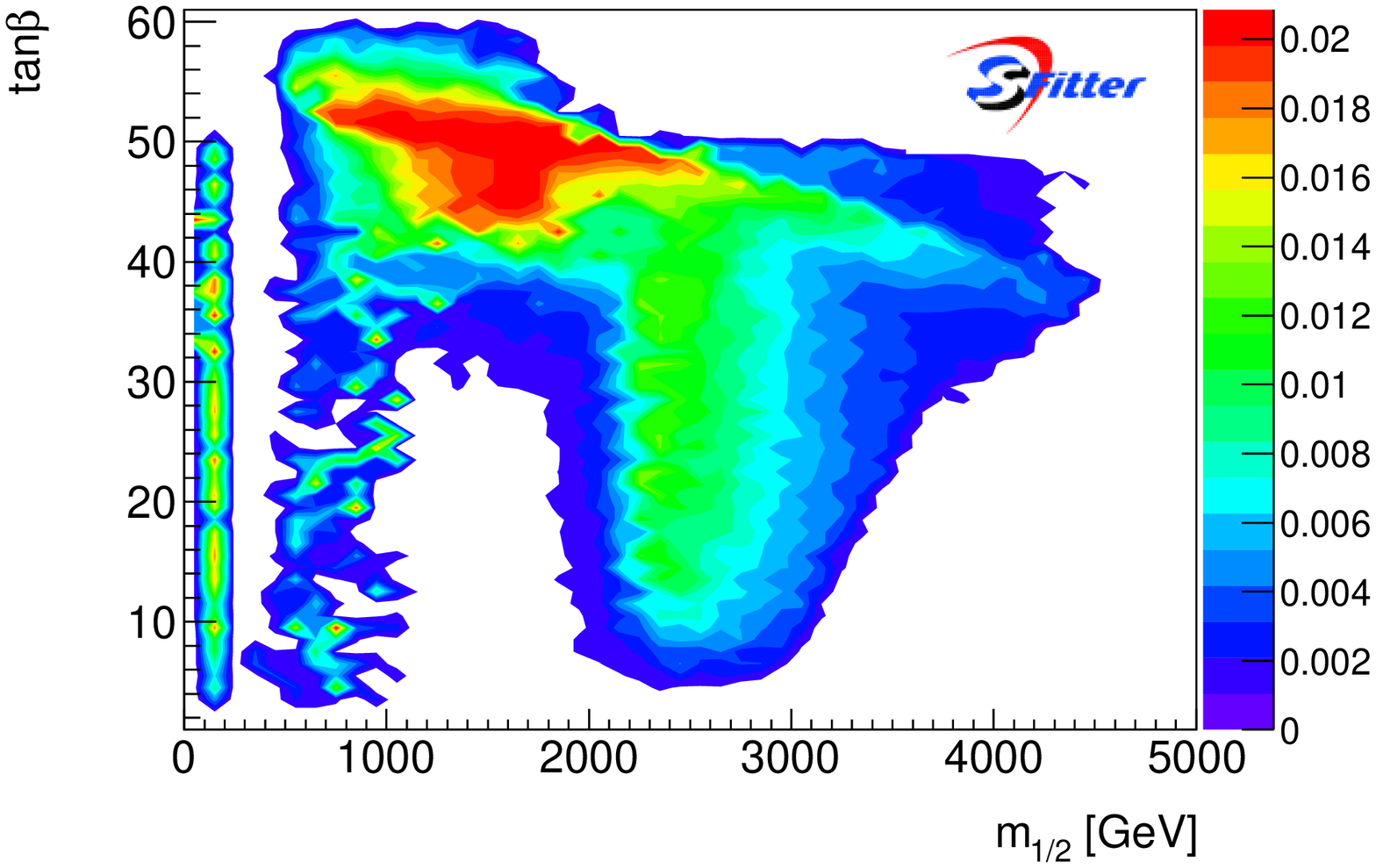}
\caption{\label{fig:mSUGRAPlanckm0m12m0A0} Profile likelihood
  projections onto the ($\Mzero,\MOneHalf$) plane, the
  ($\Mzero,\Azero$) plane, the ($\Mzero,\tanb$) plane, and the
  ($\MOneHalf,\tanb$) plane. All results are based on the Planck
  measurement and assume $\mu>0$.}
\end{figure}
%--------------------------------------------------

Because a priori the sign choice $\mu>0$ is favored by the $\Deltaamu$
measurement, we will discuss it first.  The four different
2-dimensional profile likelihoods for $\mu>0$ are shown in
Figure~\ref{fig:mSUGRAPlanckm0m12m0A0}. All of them use the recent
Planck measurement of the cold dark matter density.  The first
observation is the absence of a clear preference in the
$\Mzero$ values. In contrast, the dark matter relic density favors
three distinct regions in $\MOneHalf$, as introduced in
Section~\ref{sec:ann}:
\begin{enumerate}
\item the narrow stau co-annihilation strip with $\MOneHalf< 1$~TeV and
  $\Mzero<500$~GeV at moderate $\tanb$. The mass of the lightest
  slepton $\tilde{\tau}_1$ is very close to the LSP mass.
\item the $A$-funnel region with $\MOneHalf\approx 1.7$~TeV and
  $\tan\beta\approx 50$, where the LSP mass around 745~GeV is roughly
  half the heavy Higgs mass $m_{A,H}$ and the heavy Higgs states have
  a sizeable width to allow for a spread-out $s$-channel
  annihilation.
\item the $h$-funnel region with $\MOneHalf\approx 130$~GeV, where the
  bino-LSP mass of 60~GeV is about half the mass of the lightest
  Higgs. The dominant dark matter annihilation process is the resonant
  $s$-channel annihilation via the lightest Higgs boson. Because of
  the link between the LSP and gluino masses, this channel could
  typically be ruled out by direct LHC searches.
\end{enumerate}
Two additional well--known parameter regions~\cite{jay} are
explicitly excluded by our bounds of the parameter space. We
nevertheless confirm that they would appear in an extended parameter
scan, namely
\begin{itemize}
\item[4.] the focus point region~\cite{drees_martin,focus_point} with
  its $WW$ annihilation channel at $\Mzero\in[3,20]$~TeV, and
  $\MOneHalf\in[0.2,20]$~TeV. This region is mainly excluded by
  Xenon100~\cite{Buchmueller:2012hv}, except for a few points such as
  SPS2~\cite{snowmass} which is ruled out by LHC
  exclusions~\cite{Aad:2012fqa,hCMS}.
\item[5.] the stop co-annihilation strip with $\Azero/\Mzero\in[3,6]$
  and $\Azero/\Mzero\in[-15,-3]$.
\end{itemize}
In particular the size of the $A$-funnel region is then defined by the
light Higgs mass constraint. Relating the Higgs mass constraint we
need to be a little careful. In Section~\ref{sec:models} we have seen
that the relevant trilinear coupling $A_t$ mostly scales with
$\MOneHalf$. The main contribution to the light Higgs mass comes from
the two top squarks, so the relatively heavy Higgs mass pushes the
preferred physical stop masses to large values. According to
Eq.\eqref{eq:msugra_at} negative values of $\Azero$ will increase
$|A_t|$, leading to a larger stop mass splitting and hence a smaller
mass of the lighter stop mass eigenstate.  Indeed, we find that the
different measurements prefer $\Azero > 0$, while large negative
$\Azero$ values and low $m_0$ values are disfavored by the Higgs mass
constraint.\bigskip

In the lower panels of Figure~\ref{fig:mSUGRAPlanckm0m12m0A0} we see
that large $\tanb$ values are clearly favored, independently of
$\Mzero$.  An exception appears only for large $\Mzero$ values, where
the allowed range in $\tanb$ becomes sizeable.  The dark blue area for
$500\lesssim\Mzero\lesssim 3000$~GeV and $\tan\beta<35$ is 
disfavored by the Higgs
mass measurement. Large values of $\tan \beta$ are needed to increase
its value, while the stop masses are fairly independent of $\Mzero$.
Dark matter plays the key role in excluding the white area around
$\Mzero\approx 3.5$~TeV.

%--------------------------------------------------
\begin{table}[b!]
  \centering
  \begin{tabular}{l|c|c|c|c|c|c|c}
\hline
& $\Mzero$ & $\MOneHalf$ &  $\tan\beta$ & $\Azero$ & $m_t$ & $\chisq$/dof & $\chisq$/dof (LHCb) \\
\hline
co-annihilation &    442 &  999 & 24.6 & -1347 & 174.0 & 49.0/75 & 49.0/75 \\
$A$-funnel      &   1500 & 1700 & 46.5 &  2231 & 173.9 & 48.9/75 & 49.2/75\\
$h$-funnel      &   4232 &  135 & 26.6 & -2925 & 174.2 & 46.1/75 & 46.1/75 \\
\hline
  \end{tabular}
\caption{Illustration of best--fit parameters for the three regions of
  mSUGRA: $A$-funnel, $h$-funnel, and co-annihilation with $\mu>0$.
  The corresponding $\chisq$ is given in column 7. The last column
  illustrates the impact on the new LHCb measurement of BR($B_s \to
  \mu^+\mu^-$).}
\label{tab:mSUGRAparam}
\end{table}
%--------------------------------------------------

A similar feature, albeit a small anti-correlation, can be seen in
the $\MOneHalf$ vs $\tanb$ plane. The slight anti-correlation for
large $\tanb$ and $\MOneHalf$ is attributed to the Higgs masses.
First, the light Higgs mass increases with a larger stop mass and
hence growing $\MOneHalf$, so smaller values of $\tan \beta$ become
possible.  Moreover, the dark matter relic density can be reached
through the pseudoscalar annihilation funnel.  Because a decrease of
$\tanb$ increases the heavy Higgs masses, the dark matter relic
density forces a simultaneous increase in $\MOneHalf$ and hence the
LSP mass. This keeps the mass ratio around 2:1.\bigskip

In Table~\ref{tab:mSUGRAparam} we show the best--fit solutions in the
mSUGRA parameter space. In the last column we compare the log-likelihood
obtained with the updated measurement of BR($B_s \to \mu^+\mu^-$)
showing that the results do not depend on this observable.  The three
preferred regions with their distinct dark matter annihilation
processes are kept separate. For the $A$ and $h$ funnels in mSUGRA
the LSP has roughly the same gaugino--higgsino composition.  It is
dominantly a bino and annihilates to $b\bar{b}$ final states. In the
co-annihilation point the annihilation goes into $\tau \tau$ final
states, helped by the process $\tilde{\tau}\tilde{\chi_1^0}\rightarrow
A \tau$.  The general preference for large $\Mzero$ values from the
dark matter constraints and the lightest Higgs mass overrides the
favorite regions for $(g-2)_\mu$, which until recently dominated the
corresponding analyses.  The $\Deltaamu$ contribution to $\chisq$
becomes a constant offset.

%--------------------------------------------------
\begin{table}[t]
  \centering
  \begin{tabular}{c|c|c|c||c|c|c|c||c|c|c|c||c|c|c|c}
\hline
 & co-ann & $A$ & $h$ &
 & co-ann & $A$ & $h$ &
 & co-ann & $A$ & $h$ &
 & co-ann & $A$ & $h$ \\
\hline
$\tilde{e}_L$          &  792   &  1860 & 4210 &
$\tilde{g}$            &  2178 &   3596 & 476 &
$\tilde{q}_L$          &  2020  &   3527 & 4174 &
$h$                    & 123.0   &  123.0 & 124.8\\
$\tilde{e}_R$          &  575   &  1621 & 4223 &
$\tilde{\chi}_1^0$     &  429  &    745 &  59 &
$\tilde{q}_R$          &  1939  &   3397 & 4192 &
$H$                    & 1423   &  1498 & 3624\\ 
$\tilde{\nu}_{eL}$      &  788  &  1858 & 4209 &
$\tilde{\chi}_2^0$     &  809  &   1379 &  118 &
$\tilde{b}_1$          &  1754  &   3046 & 3190 &
$A$                    & 1423   &  1498 & 3624\\
$\tilde{\mu}_L$        &  792  &  1860  & 4210 &
$\tilde{\chi}_3^0$     &  -1407  &  -1588 & -507 &
$\tilde{b}_2$          &  1849  &   3101  & 3877 &
$H^+$                  & 1425  &  1500 & 3625 \\ \cline{13-16}
$\tilde{\mu}_R$        &  575  &  1621  & 4223 &
$\tilde{\chi}_4^0$     &  1412 &   1603 & 512 &
$\tilde{t}_1$          &  1426  &   2771 & 2374 \\
$\tilde{\nu}_{\mu L}$  &  788  &  1858  & 4209 &
$\tilde{\chi}_1^+$     &  810  &   1379 &  119& 
$\tilde{t}_2$          &  1791  &   3105 & 3212 \\ \cline{9-12}
$\tilde{\tau}_1^-$     &  430  &  1103 & 3920 &
$\tilde{\chi}_2^+$     &  1412 &   1603  & 514\\ \cline{5-8}
$\tilde{\tau}_2^-$     &  756 &  1666  & 4062 \\
$\tilde{\nu}_{\tau L}$ &  744 &  1661  & 4061 \\ \cline{1-4}
  \end{tabular}
\caption{\label{table:massesMSUGRA} Supersymmetric particles' masses
  (in GeV) for the three best--fit points shown in
  Table~\ref{tab:mSUGRAparam}. They correspond to the favored regions:
  $A$-funnel, $h$-funnel, and co-annihilation with $\mu>0$.}
\end{table}
%--------------------------------------------------

The influence of the top mass and its uncertainty cannot be neglected,
as we see for example in the $h$-funnel region. Compared to the
nominal value of 173.5~GeV in Table~\ref{tab:data} the best fit result
shown in Table~\ref{tab:mSUGRAparam} is increased by 0.7~GeV. This
increase leads to a slight reduction of $\Mone$ by at most 0.1~GeV and
an increase of $\mu$ from 350~GeV to 490~GeV. For the LSP this implies
a larger mass by about 0.8~GeV and a decreased higgsino component by 
almost 50\%. In parallel, the Higgs mass increases by 0.2~GeV, as compared 
to
the prediction using the nominal top mass. Combining the two mass
shifts and the decreased LSP coupling to the Higgs leads to the
correct value of $\omegacdm$.\bigskip
%--------------------------------------------------
\begin{figure}[b!]
\includegraphics[width=0.49\textwidth]{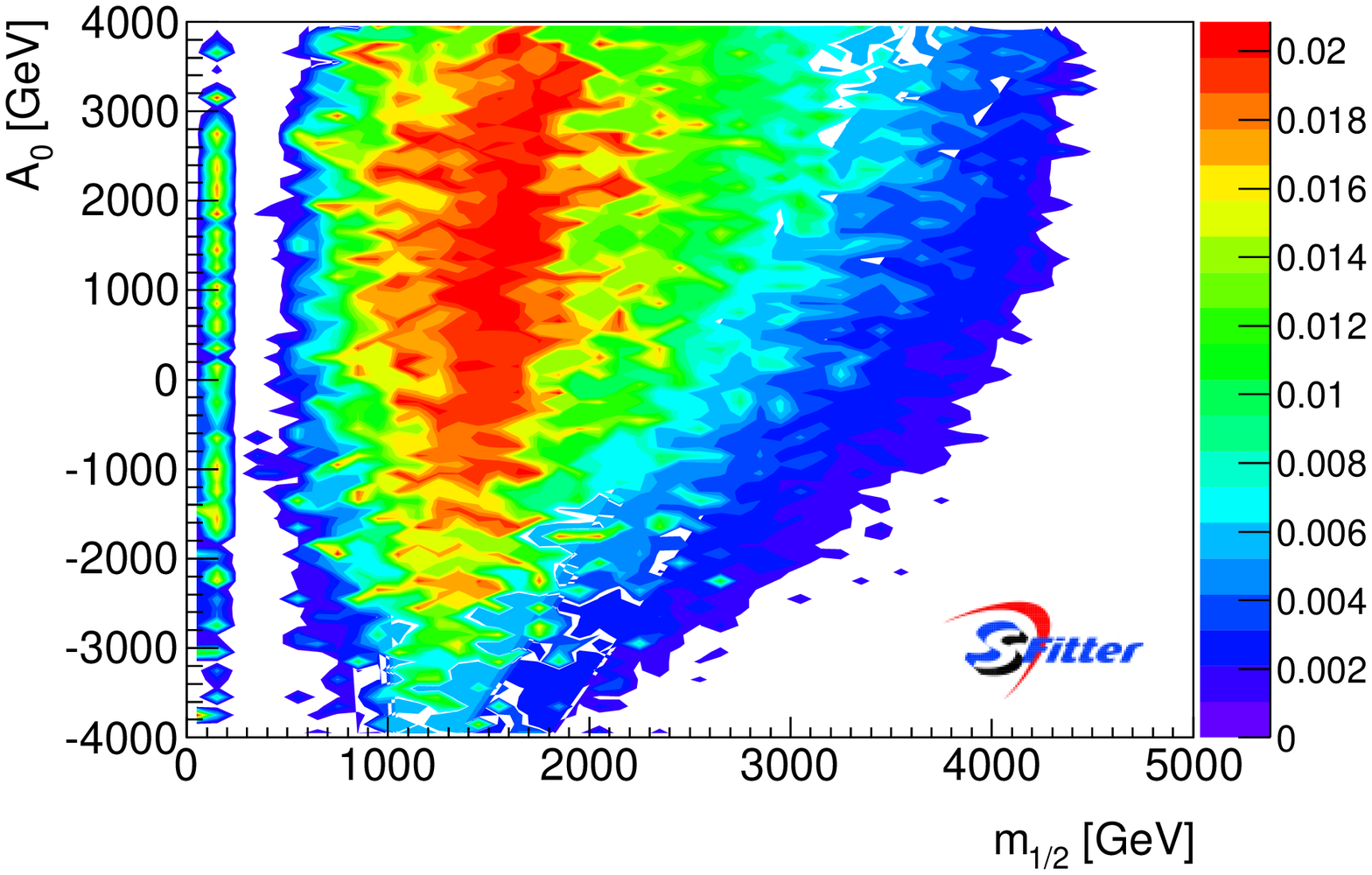}
\includegraphics[width=0.49\textwidth]{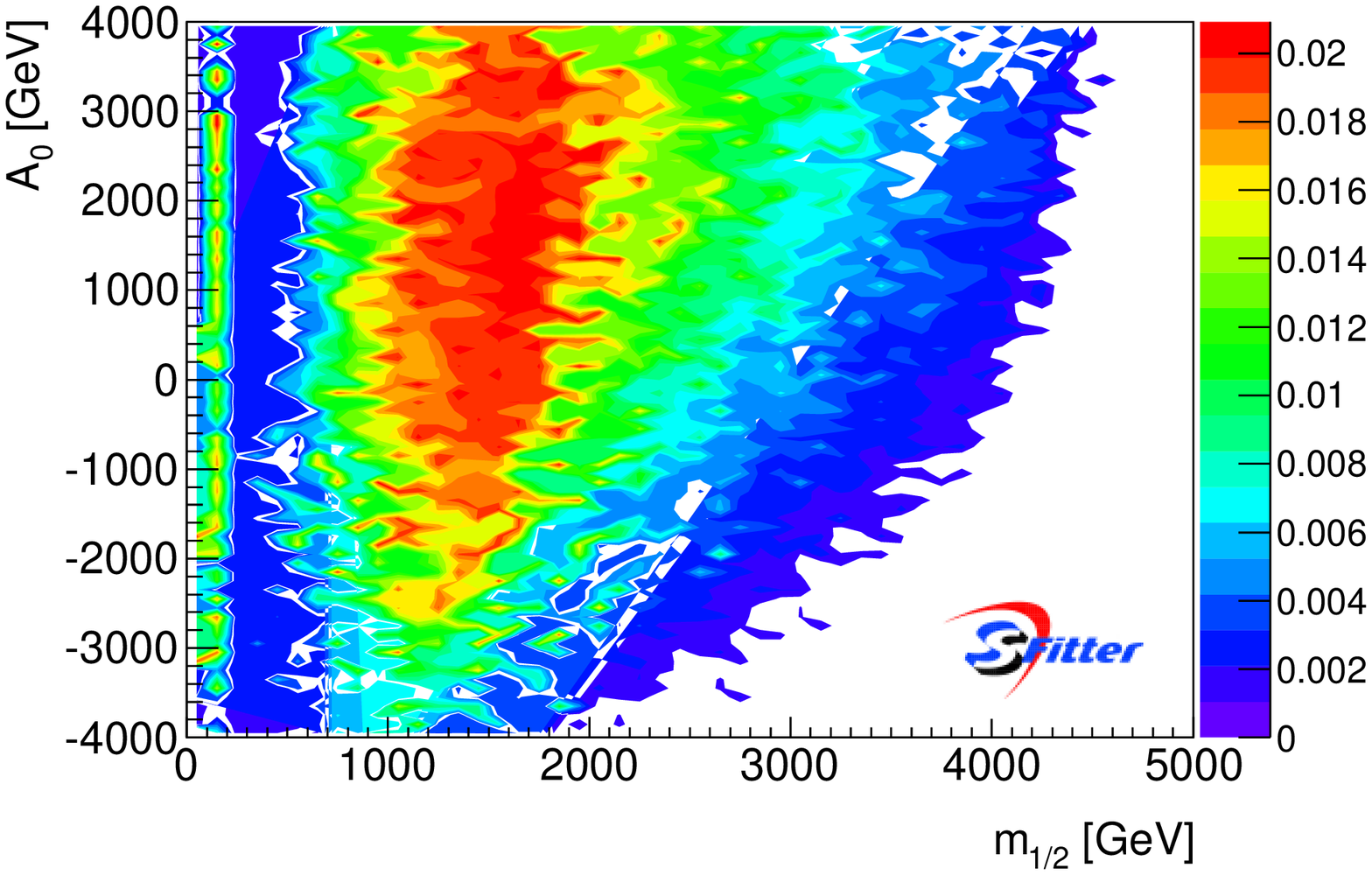} 
\caption{\label{fig:mSUGRAm12A0} Profile likelihood projection onto
  the ($\MOneHalf,\Azero$) plane using the Planck (left) and WMAP
  (right) measurements.} 
\end{figure}
%--------------------------------------------------

The complete mass spectrum of the sparticles corresponding to the three
points is given in Table~\ref{table:massesMSUGRA}. The $h$-funnel has
a relatively light gluino of 476~GeV, driven by the low LSP mass. The
squark masses turn out heavy. Because the available mSUGRA limits from
ATLAS~\cite{Aad:2012fqa} and CMS~\cite{hCMS} are calculated for
different values of $\Azero=0$ and $\tanb=10$ the results cannot be
applied directly, but it is clear that this parameter point
will be excluded by inclusive squark and gluino searches at the
LHC. The only obvious way to hide light gluinos in these analyses
would be to complement them with mass-generate squarks, such that the
decay jets become too soft to be observed~\cite{jamie}. However, in
mSUGRA the squark masses are linked to the stop masses, and light stop
masses are ruled out by the Higgs mass constraint. Hence, for mSUGRA
the list of non-excluded dark matter annihilation channels given in
Section~\ref{sec:ann} is reduced to stau co-annihilation and the
$A$-funnel within the parameter space considered in this analysis.\bigskip

As mentioned above, one of the key motivations of this analysis is to
see the impact of the recent Planck measurements, in comparison to the
WMAP-9year results.  The most visible difference can be observed in
the ($\MOneHalf$, $\Azero$) plane in Figure~\ref{fig:mSUGRAm12A0}.
The general features are very similar. In addition, the
separation between the light Higgs funnel region and the rest of the
plane becomes clearer with the new and improved Planck
measurement. This reflects the essentially equivalent central values
but smaller error bars on $\omegacdm$.

%%%%%%%%%%%%%%%%%%%%%%%%%%%%%%%%%%%%%%%%%%%%%%%%%%%%%%%%%%%%%%%%%%%%%%%%%%%%%%%%
\subsection{Bayesian probability for positive $\mu$}

%--------------------------------------------------
\begin{figure}[b!]
\includegraphics[width=0.49\textwidth]{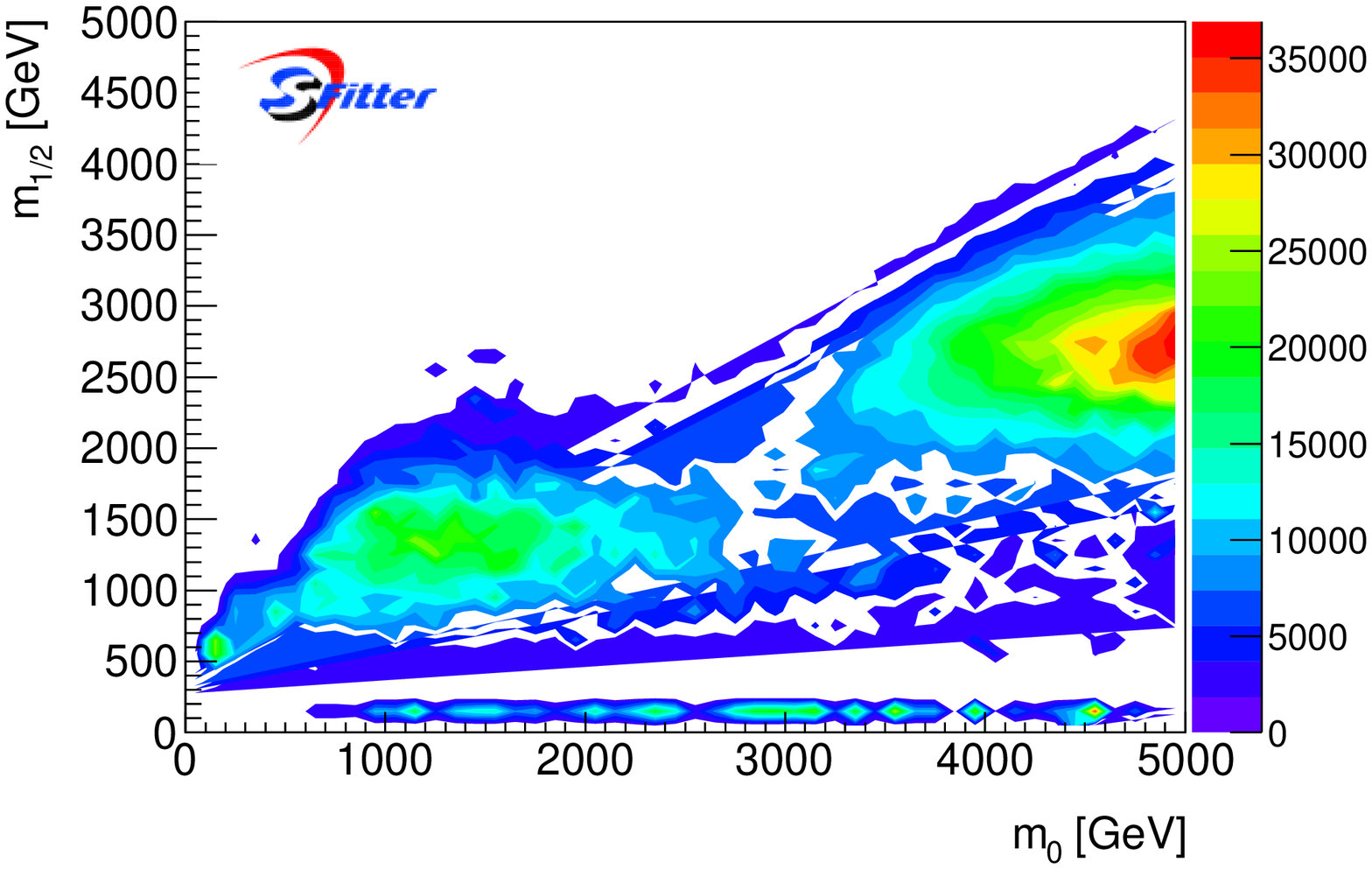}
\includegraphics[width=0.49\textwidth]{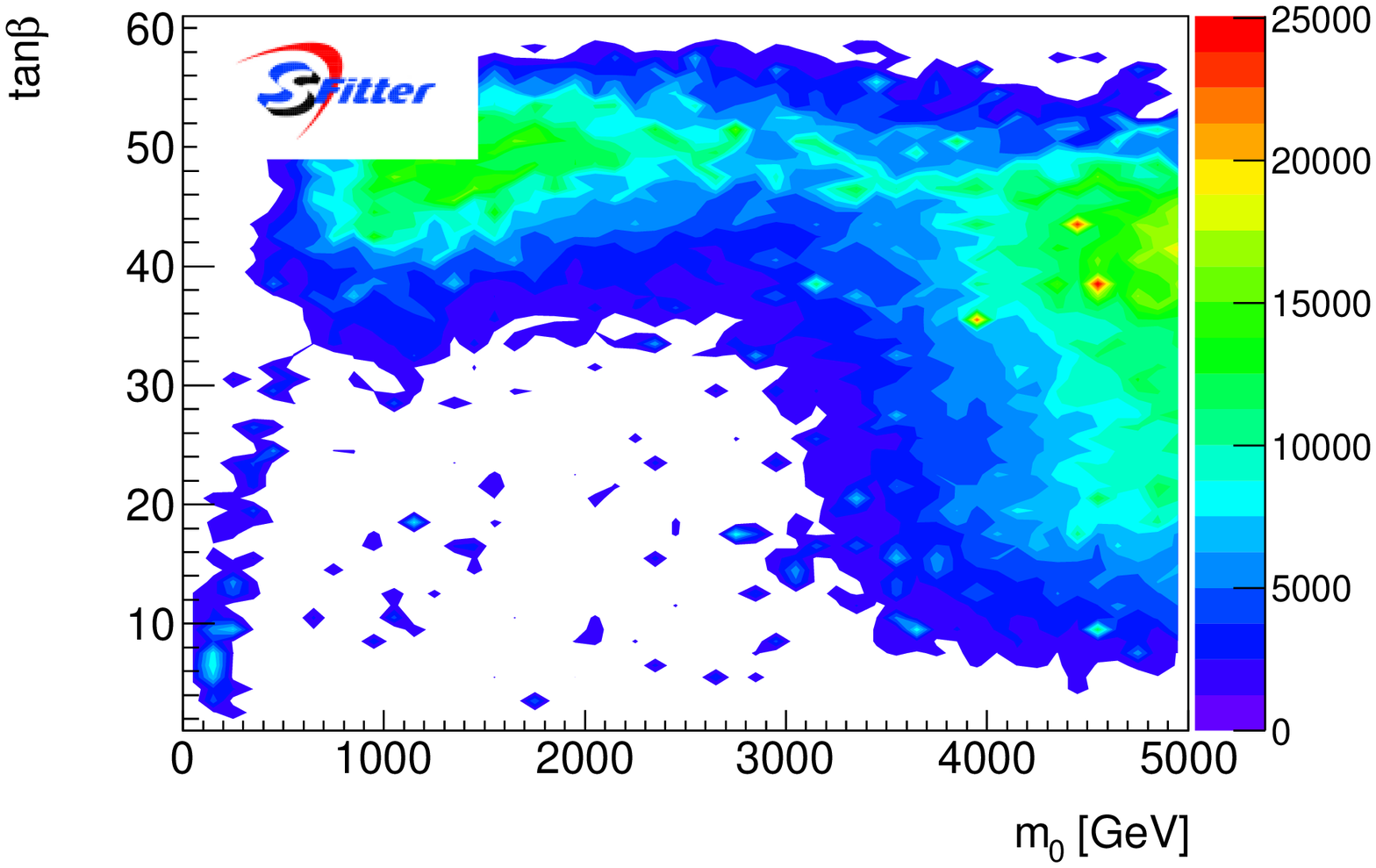}
\includegraphics[width=0.49\textwidth]{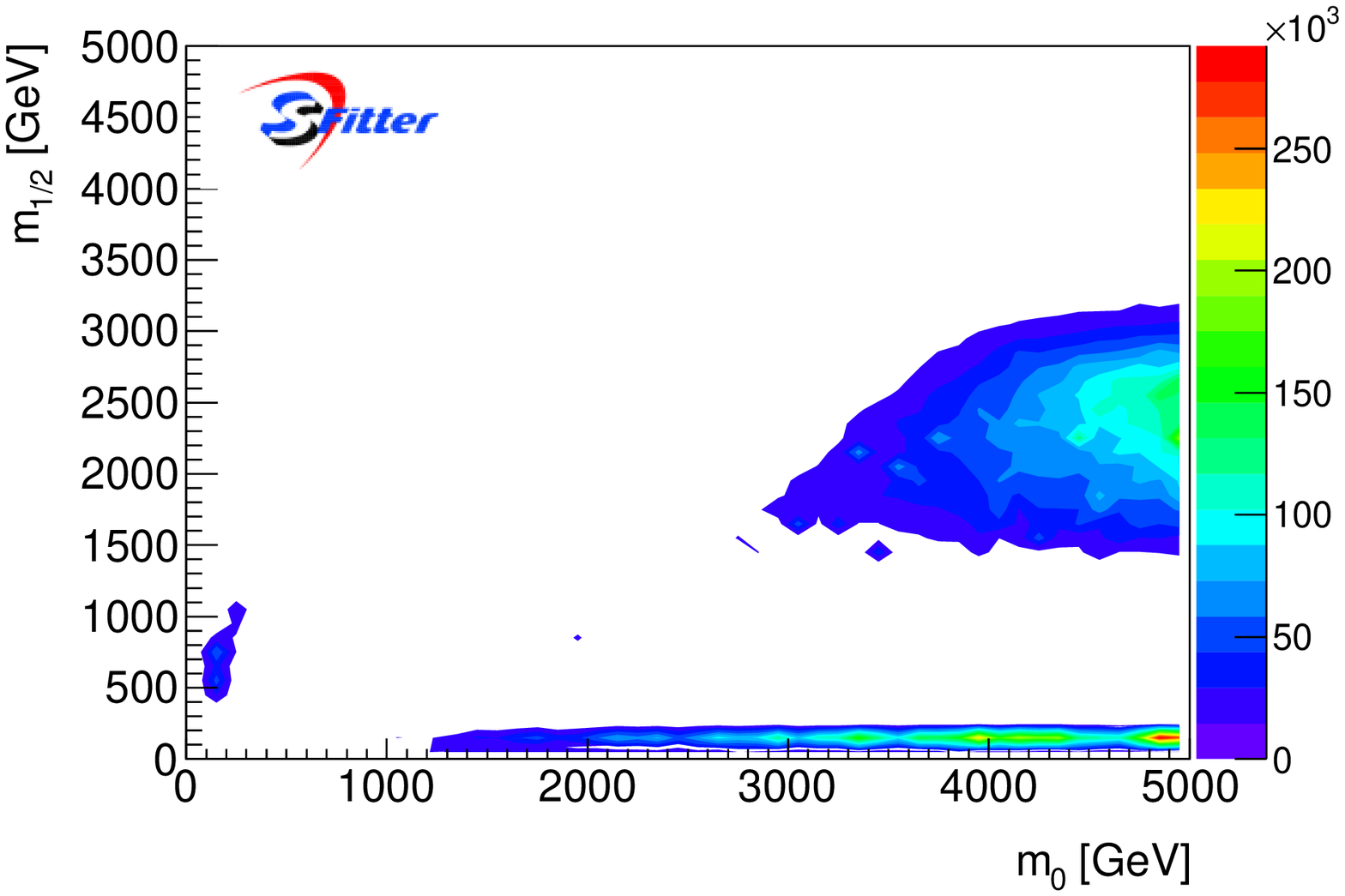}
\includegraphics[width=0.49\textwidth]{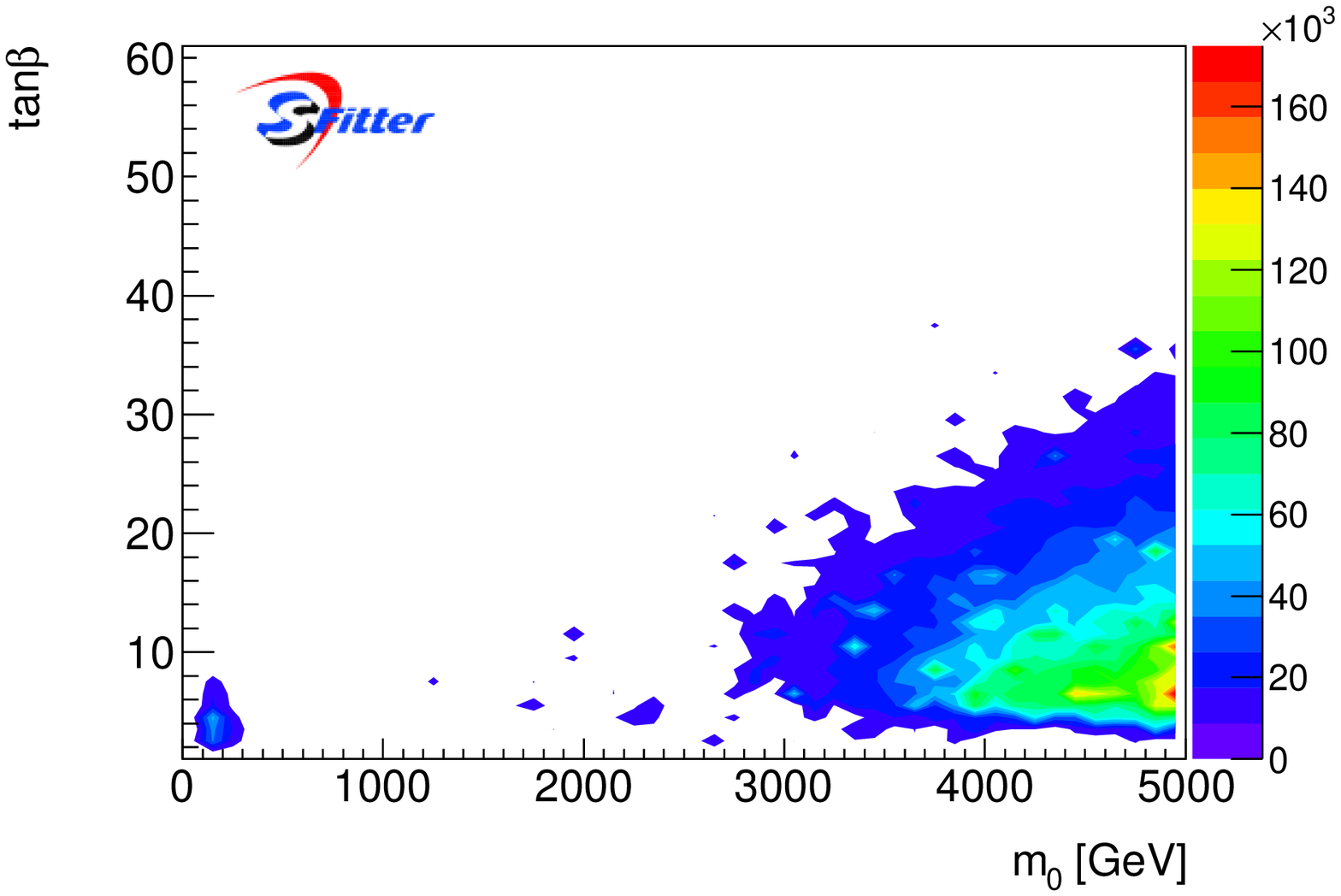}
\caption{\label{fig:mSUGRAm0m12m0tanbBayes} Bayesian projection onto
  the ($\Mzero,\MOneHalf$) plane (left) and the ($\Mzero,\tanb$) plane
  (right) for a $\tanb$--flat prior (top) and a high--scale flat prior
  (bottom). All results are based on the Planck measurement and assume
  $\mu>0$.}
\end{figure}
%--------------------------------------------------

To this point we have only relied on profile likelihood
projections. While Frequentist and Bayesian approaches cannot be
expected to give equivalent answers (because they ask different
questions) they can still give complementary information. In
Figure~\ref{fig:mSUGRAm0m12m0tanbBayes} we show the Bayesian
projections onto the ($\Mzero,\MOneHalf$) and ($\Mzero,\tanb$) planes,
using the consistent $\tanb$--flat and high--scale flat priors
discussed in Section~\ref{sec:models}.

Both, the ($\Mzero,\MOneHalf$) plane and the ($\Mzero,\tanb$) plane
using the consistent high--scale flat prior show similar features as
for the profile likelihood approach.  First, there is the well
separated low--$\MOneHalf$ solution from the light Higgs
funnel. Second, the narrow co-annihilation strip is hard to see, but
still present. Finally, the $A$-funnel bulk region is divided in low
and high $\Mzero$ values and shows a clear preference for $\Mzero >
4.5$~TeV and $\MOneHalf \approx 2.8$~TeV. This can be explained by
the volume effect when integrating over $\tan \beta$, $A_0$, and $m_t$: 
the best-fit value around $\MOneHalf=1.5$~TeV has a low
probability for most $\tanb$ values, except for $\tanb=40 - 50$. In
contrast, for $\MOneHalf \approx 2.5$~TeV the preferred region extends
over almost all $\tan \beta$ values. In general, $m_t$ moves
significantly below its nominal value to accommodate the $A$-funnel
region, but covering a larger range for large $\Mzero$.  All of these
features can also be seen in the profile likelihood analysis, but
they only develop two well defined preferred regions after we
integrate the Bayesian probabilities.

The whole picture changes significantly when we instead use a
low--energy prior, flat in $\tan \beta$, in the Bayesian analysis. In
the ($\Mzero,\MOneHalf$) plane the low-$\Mzero$ part of the bulk
solution vanishes.  In the ($\Mzero,\tanb$) plane, suddenly low
$\tanb$ values are favored. This is simply an effect of the relative
difference in priors shown in Eq.\eqref{eq:jacobian}. Such a prior
dependence suggests that our information is not yet sufficient to draw
conclusions on Bayesian favored regions.

%--------------------------------------------------
\begin{figure}[t]
\includegraphics[width=0.49\textwidth]{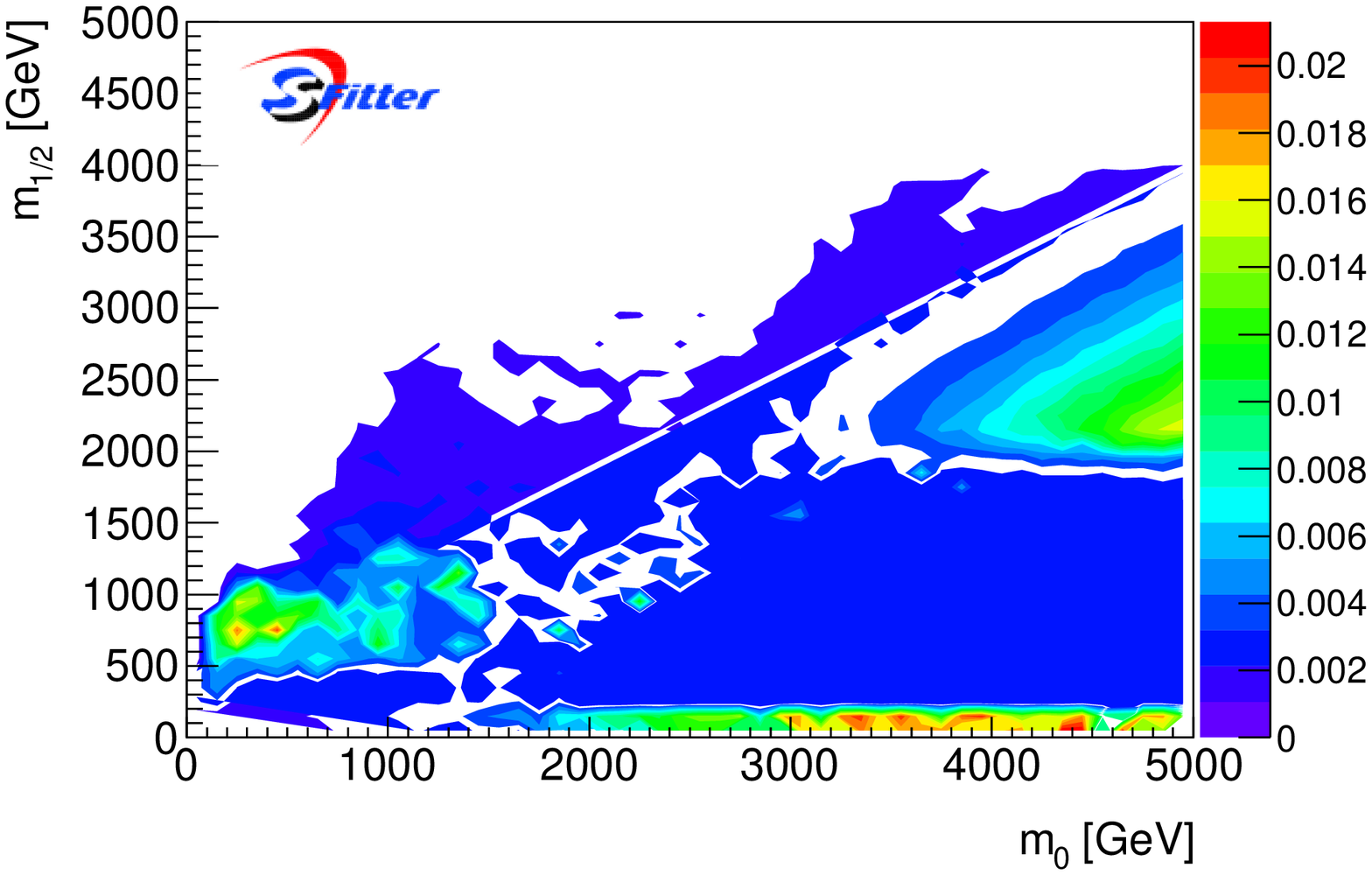}
\includegraphics[width=0.49\textwidth]{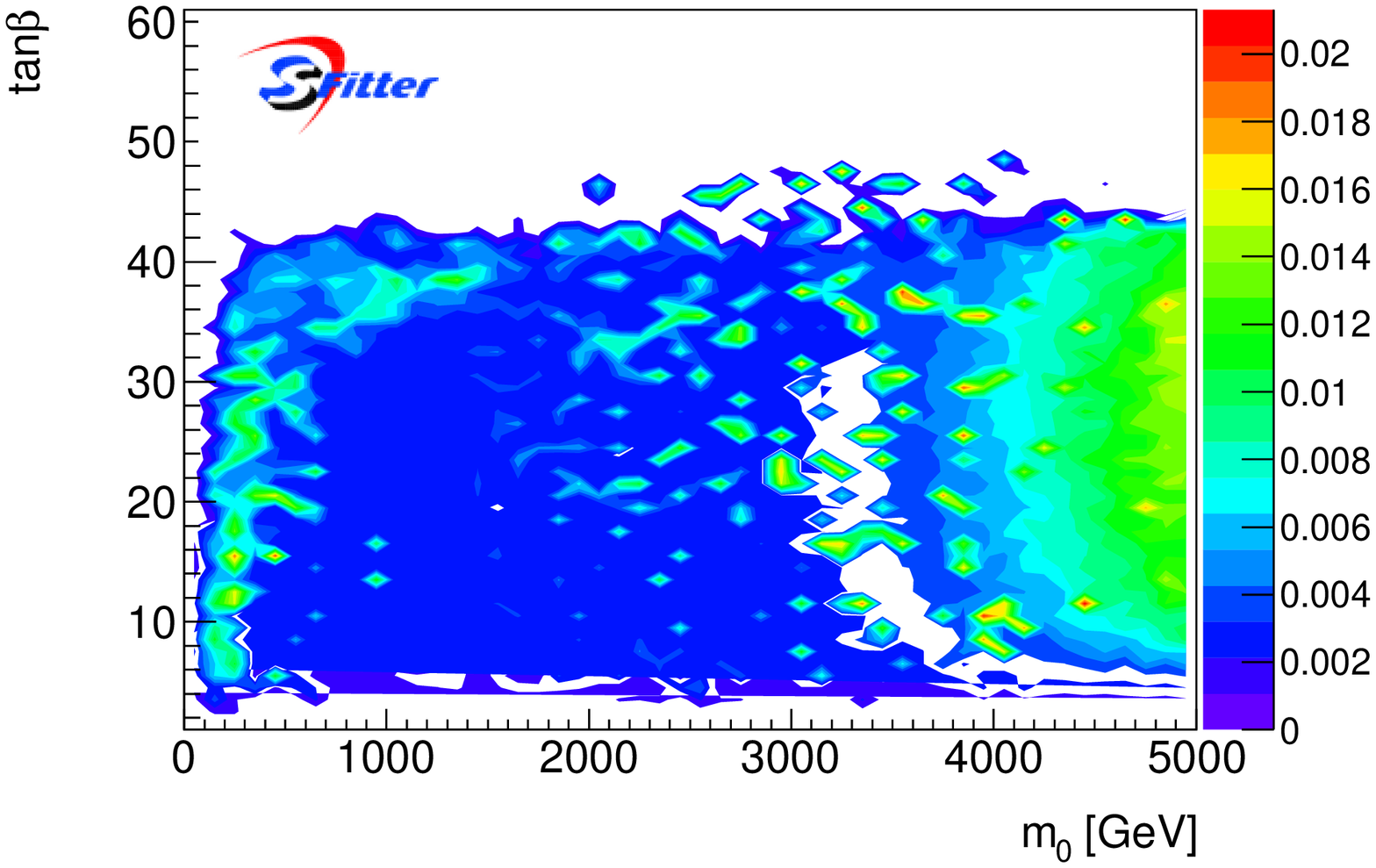}
\caption{\label{fig:mSUGRAminusm12A0} Profile likelihood projections
  onto the ($\Mzero,\MOneHalf$) and ($\Mzero,\tanb$) planes. All
  results are based on the Planck measurement and assume $\mu<0$.}
\end{figure}
%--------------------------------------------------

%%%%%%%%%%%%%%%%%%%%%%%%%%%%%%%%%%%%%%%%%%%%%%%%%%%%%%%%%%%%%%%%%%%%%%%%%%%%%%%%
\subsection{Negative $\mu$}

Finally, we turn to $\mu<0$. From the argument above we would expect
similarly good fits with a finite log-likelihood offset from
$\Deltaamu$.  In Figure~\ref{fig:mSUGRAminusm12A0} we indeed observe
similar features as for $\mu>0$, but on the absolute scale of the
log-likelihood only the $h$-funnel region at low $\MOneHalf$ retains
its features. The $A$-funnel region at $\MOneHalf\approx 1.5$~TeV is
now clearly disfavored.

The correlation in $\tanb$ vs $\Mzero$ sheds some light on this
feature: for large values of $\tanb$ and $\mu<0$ values the
cancellation in the off--diagonal entries of the third generation
squark mass matrices fails. This will lead to light sbottoms and stops
with very large couplings to the heavy Higgs states.  They will
trigger conflicts with heavy flavor measurements and eventually with
the perturbativity of the renormalization group equations.  The best
solutions for $\mu<0$ are hence restricted to the light Higgs funnel
and the co-annihilation regions at low values of $\MOneHalf$.

%%%%%%%%%%%%%%%%%%%%%%%%%%%%%%%%%%%%%%%%%%%%%%%%%%%%%%%%%%%%%%%%%%%%%%%%%%%%%%%%
\section{MSSM analysis}
\label{sec:mssm}

Going from a strongly constrained model such as mSUGRA to the MSSM increases
the number of free parameters. The ultimate goal of such an analysis is to shed
light, with enough experimental constraints, on which scenarios of SUSY
breaking are favored.  We choose to constrain 13 parameters plus the top
mass. Our parameter space is bounded by $\tanb<61$, $(\Mone,\Mtwo)<4$~TeV,
$(M_{\tilde{\mu}_{L/R}},M_{\tilde{\tau}_{L/R}},M_{\tilde{q}_{3L}},M_{\tilde{t}_R})<5$~TeV,
$(|A_{\tau}|,|A_t|)<4$~TeV, $m_A<5$~TeV and $|\mu|<2$~TeV. 
This number is considerably larger than the number of strong
constraints or measurements we apply in our analysis, rendering the
analysis quite complex in terms of likelihood maximization. On the other
hand, now, different sub-sectors of parameters largely decouple.
We
analyze the MSSM parameter space with 100 Markov chains of 200000
points each, leading to a total number of $2\times 10^7$ of tested
samples. For the
convergence parameter $\max[\hat{R}]$ typical values are 1.005 and
better. \bigskip

The measured light Higgs mass essentially depends on three parameters:
the heavy Higgs mass scale $m_A$, which has to be large to accommodate
the 126~GeV measurement; $\tan \beta$ which has to be large enough to
not delay the decoupling regime in $m_A$; and finally the geometric
mean of the two stop masses $\sqrt{m_{\tilde{t}_1} m_{\tilde{t}_2}}$,
which again has to be large. In terms of MSSM parameters the latter
needs to be computed from the three entries in the stop mass matrix,
including $A_t$. The stop masses are the key parameters, but are neither
strongly related to the dark matter sector nor to the light--flavor
squark--gluino mass plane.  In addition, they are directly linked to
the solution of the hierarchy problem and hence to the motivation of
supersymmetry.\bigskip

%--------------------------------------------------
\begin{figure}[b!]
  \includegraphics[width=0.49\textwidth]{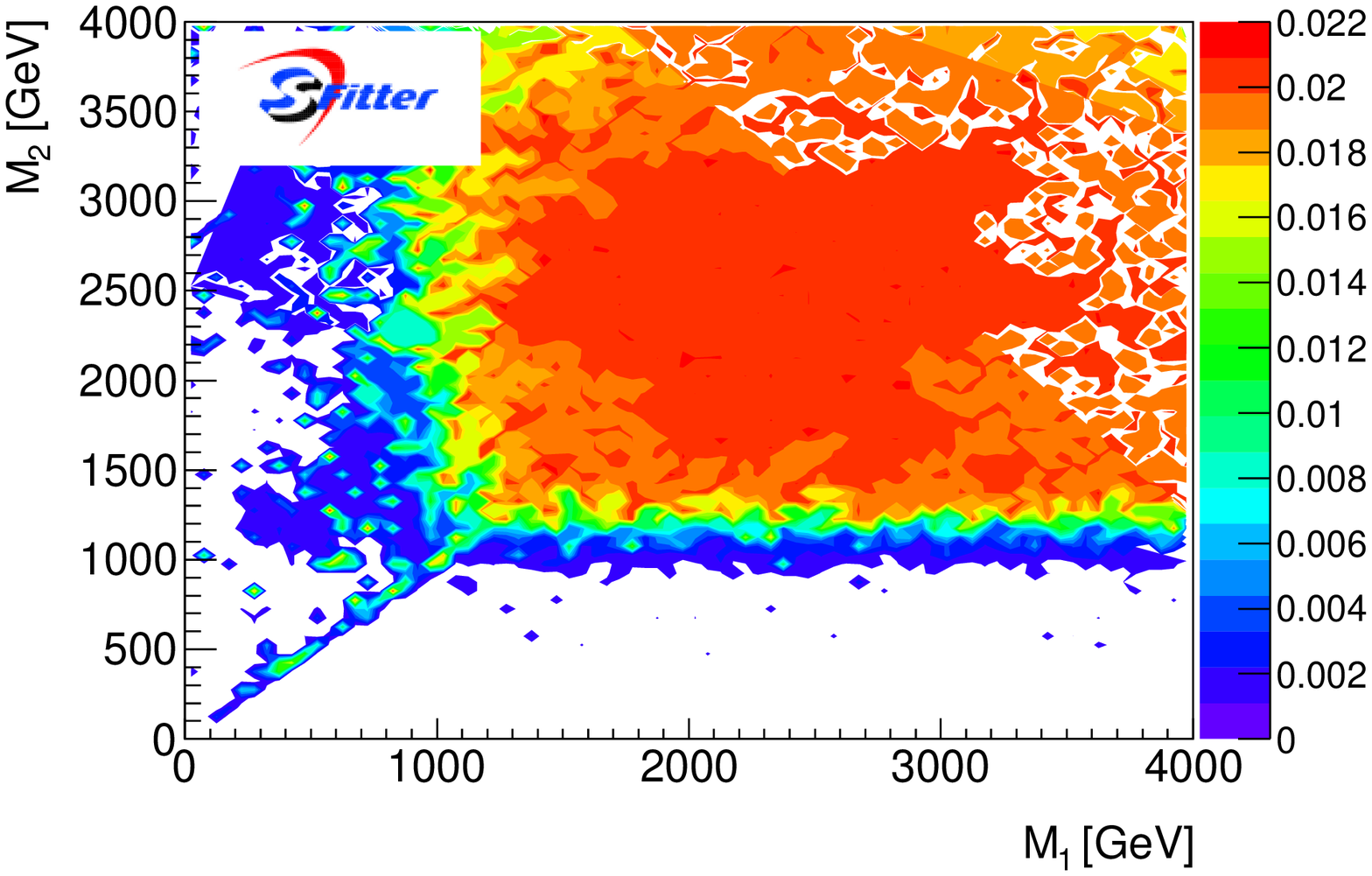}
  \includegraphics[width=0.49\textwidth]{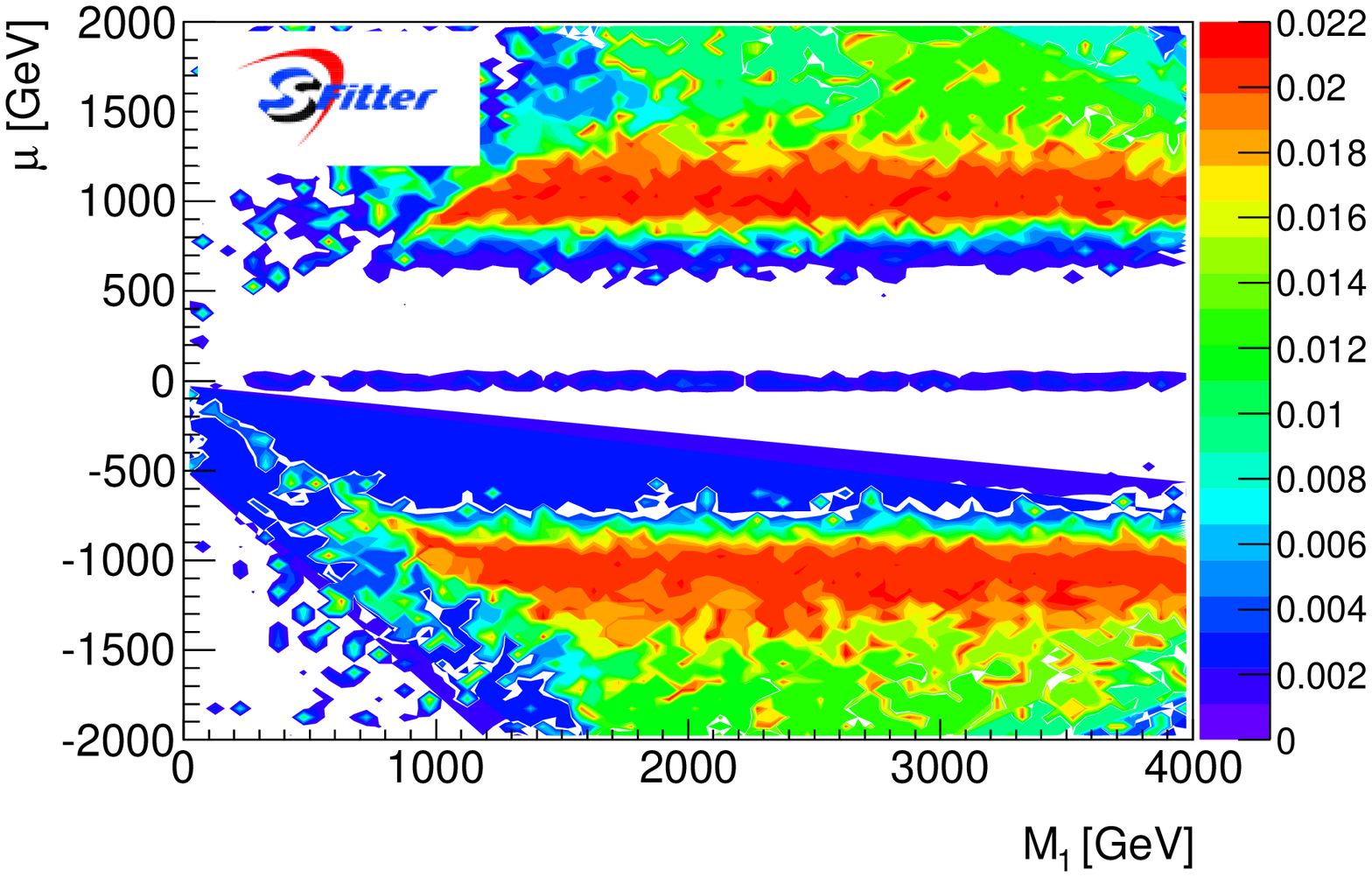} 
  \caption{\label{fig:MSSMm1m2m1muPlanck} Profile likelihood
    projection onto the ($\Mone,\Mtwo$) plane (left) and the
    ($\Mone,\Mtwo$) plane (right) for the Planck measurements.}
\end{figure}
%--------------------------------------------------

The light--flavor squark masses and the gluino mass are experimentally
constrained by searches for jet plus missing energy in LHC
experiments. While it is entirely possible to avoid these limits in
certain decay setups, the strongly interacting supersymmetric masses
are likely to lie in the several-TeV range. This tendency towards a
heavy strongly interacting SUSY sector is in line with the stop mass
constraint from the Higgs sector.\bigskip

The dark matter sector is most strongly constrained by our requirement
that the entire relic density is due to the LSP, in our case the
lightest neutralino. The neutralino masses and couplings depend on the
four parameters $\Mone$, $\Mtwo$, $\tanb$ and $\mu$. The link between
the dark matter sector and other sectors rests on the different LSP
annihilation channels, as explained in detail in the mSUGRA
section. For a sufficiently fast LSP annihilation we cannot rely on
generic scattering processes, for example with a $t$-channel slepton,
squark, or chargino. Instead, the easiest ways to reach the observed
$\omegacdm$ values are light and heavy Higgs funnels and
co-annihilation.

In general, the range of $\mu$ is strongly limited as the light
charginos and neutralinos are constrained by direct LEP searches and
$Z$~pole measurements~\cite{ALEPH:2005ab}. This results in
log-likelihood
values about ten times worse than the minimum. For example, for $\mu=
20$~GeV and variable $\Mtwo$ the typical $Z$ width is increased by
30~MeV, a large amount compared to the error of 3~MeV and hence ruled
out.\bigskip

In Figure~\ref{fig:MSSMm1m2m1muPlanck} we show the profile likelihoods
in the neutralino and chargino sector $M_1$, $M_2$, and $\mu$ for the
Planck measurement. All measurements discussed in
Section~\ref{sec:setup} are included. The log-likelihood map favors five
regions, three of which directly correspond to the mSUGRA case:
\begin{enumerate}
\item the stau co-annihilation strip diagonal in $\Mone$ vs $\Mtwo$ at
  relatively small values. Here, the mass of the lightest slepton
  $\tilde{\tau}_1$ is very close to the LSP mass.
\item the $A$-funnel region where the LSP mass is about half the heavy
  Higgs mass. This MSSM region behaves the same way as discussed for
  the simpler mSUGRA model. In Figure~\ref{fig:MSSMm1m2m1muPlanck} it
  contributes to the bulk region of the $\Mone$ vs $\Mtwo$ plane as
  well as to the correlated patterns in the $\Mone$ vs $\mu$ plane.
\item the $h$-funnel region at low $\Mone \sim 63$~GeV almost
  independent of $\Mtwo$. Unlike for mSUGRA the gluino mass is now an
  independent parameter, so the direct LHC searches decouple from the
  dark matter sector. Because the corresponding MSSM parameter space
  is tiny, the funnel appears only as distinct sets of points in
  Figure~\ref{fig:MSSMm1m2m1muPlanck}. We have checked that it
  actually is a narrow line.
\item a bino-higgsino region which appears as a strip in the $\Mone$
  vs $\mu$ plane for $\mu<0$ and $|\Mone| \approx |\mu|$. The dark
  matter annihilation proceeds through different neutral and charged
  Higgs--mediated channels, including chargino co-annihilation and
  dominantly third--generation quarks in the final state. The latter
  includes the $b\bar{b}$ final state from the $A$-funnel.
\item a large higgsino region with $\Mone,\Mtwo > 1.2$~TeV, split in
  two almost symmetric solutions $\mu\approx \pm 1.2$~TeV.  Because
  the LSP characteristics in the two regions are very similar we will
  only refer to $\mu>0$. Chargino co-annihilation dominates the
  prediction of the relic density with first and second generation
  quarks in the final state.
\end{enumerate}
As for the mSUGRA case an additional stop co-annihilation region
exists, but is not covered by our parameter range.\bigskip

%--------------------------------------------------
\begin{table}[t]
\begin{center}
  \begin{tabular}{l|c|c|c|c|c}
\hline
                       & co-ann & $A$-funnel & $h$-funnel  & bino--higgs  & higgsino  \\
\hline
$\tan\beta$         	&  25     &       18&  26.6    &       54  &          29  \\
$M_1$       		& 430     &      400&  59      &       800 &        1543  \\
$M_2$  	 		& 788  &     1500 &  960        &       2174 &        2898  \\
$\mu$ 			& 1400     &      750 & 484      &     -800 &        1070  \\
\hline
$M_{\tilde{\mu}_L}$ 	& 791      &     1586 & 4210     &      3994 &        2884 \\
$M_{\tilde{\mu}_R}$     &  573    &     2789 & 4223      &      1002 &        2790  \\
$M_{\tilde{\tau}_L}$    & 747     &     1067 & 4062      &      3744 &        3355  \\
$M_{\tilde{\tau}_R}$	 &  440     &     2789  & 3921     &      2040 &        2058\\
$A_{\tau}$ 		& -1690   &    -3038&  -2570   &      2338 &       -3533  \\
\hline
$M_{\tilde{q}_{3L}}$ 	& 1744     &     3938 & 3162     &      1683 &        2210 \\
$M_{\tilde{t}_R}$ 	& 1441   &     3997 & 2319       &      2111 &        2984  \\
$A_t$ 			& -2142    &    -3158 & -1230     &     -2162 &      -3026  \\
\hline
$m_A$ 			& 1423    &      781 & 3626       &       1000 &         784\\
\hline
$\Mtop$ 		& 174.0  &    173.5 & 173.5      &     173.6 &       173.5  \\
\hline
$\chisq$/dof             & 47.9/65 & 44.2/65  & 46.5/65 &   42.5/65 &     37.8/65  \\
\hline
  \end{tabular}
  \hspace*{0.5cm}
  \begin{tabular}{l|c|c|c|c|c}
\hline 
                        & co-ann & $A$-funnel & $h$-funnel & bino--higgs & higgsino  \\
\hline		       					       
$\tilde{\chi}_1^0$     & 429   &   398  & 58.5        &  768 & 1066    \\
$\tilde{\chi}_2^0$      & 783   &   749 & 480        & -801 & -1071     \\
$\tilde{\chi}_3^0$     & -1402  &  -751  & -488          & 829 & 1545   \\
$\tilde{\chi}_4^0$     & 1406  &  1506 & 969          &  2178 & 2900    \\
$\tilde{\chi}_1^+$     & 784  &   747 & 480          &  799 & 1069    \\
$\tilde{\chi}_2^+$     & 1407  &  1506  & 969         & 2178 & 2900    \\
\hline
$h$                    & 123.2 &  125.3 & 122.1     &  123.2 & 124.5   \\
$H$                    & 1423  &   781  & 3626       &  1000 & 784       \\ 
$A$                    & 1423   &   781 & 3626      &  1000 & 784      \\
$H^+$                  & 1425   &   785 & 3627       &  1003 & 788      \\
\hline
  \end{tabular}
\end{center}
\caption{Left: examples of best--fit points for the MSSM are shown
  together with $\chisq$ per degrees of freedom.  Right: neutralino
  and chargino masses for the best--fit MSSM points.  The masses are
  given in GeV. }
\label{table:mMSSMparam}
\end{table}
%--------------------------------------------------

%--------------------------------------------------
\begin{figure}[b!]
  \includegraphics[width=0.49\textwidth]{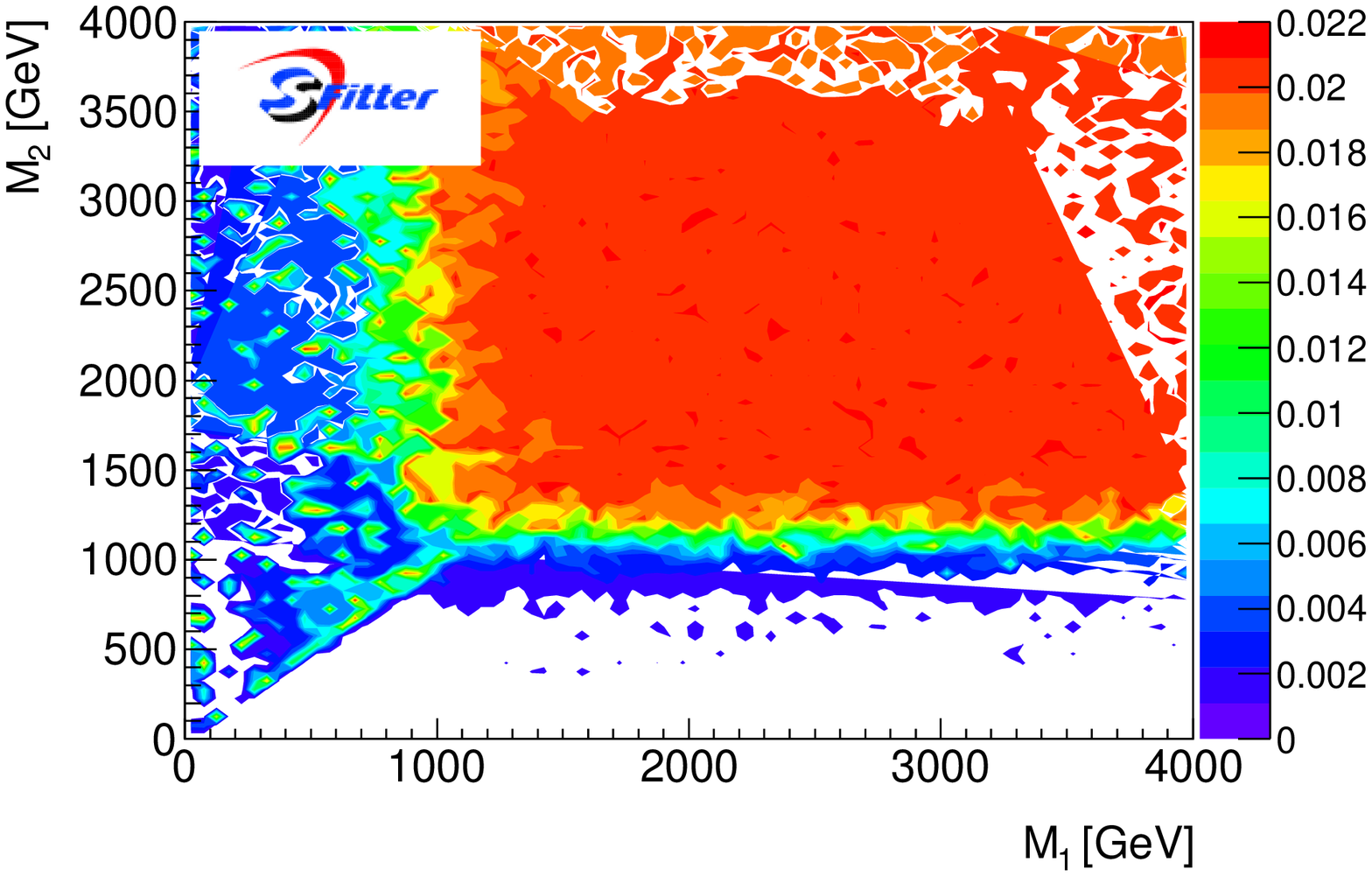}
  \includegraphics[width=0.49\textwidth]{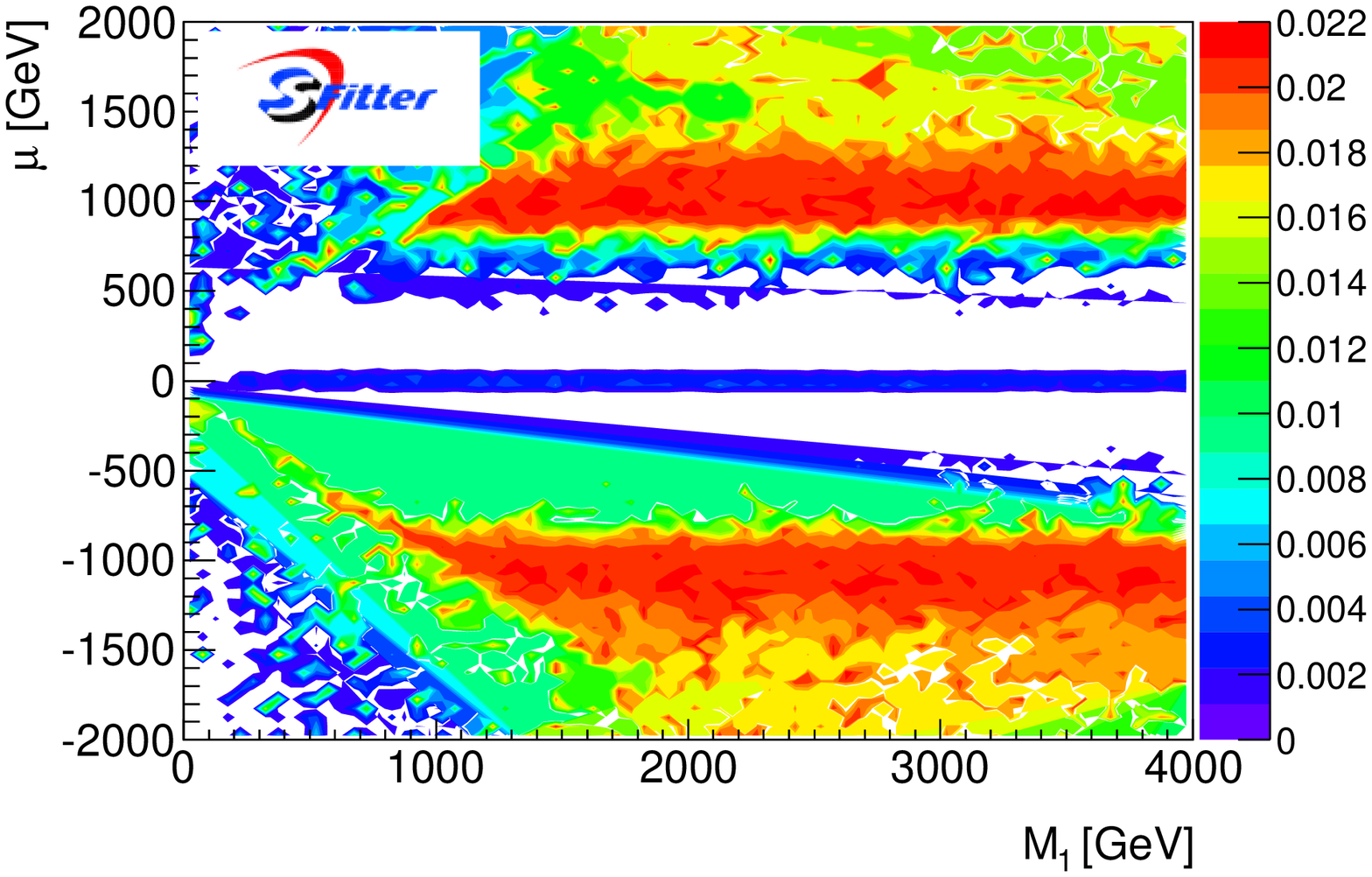}
  \caption{\label{fig:MSSMm1m2m1muWmap} Profile likelihood projection
    onto the ($\Mone,\mu$) plane (left) and the ($\Mone,\Mtwo$) plane
    (right) for the WMAP results.}
\end{figure}
%--------------------------------------------------

In Table~\ref{table:mMSSMparam} we give examples for individual
best-fitting parameter points in each of these regions.  As the
parameters are less correlated in the MSSM than in  mSUGRA,
the top quark mass parameter essentially does not move from
its measured value.
None of these points are excluded from LHC direct SUSY Higgs searches
such as~\cite{HiggsSUSY}.
For the bulk of the solutions the hierarchy in the neutralino sector
favors a smaller $\mu$, corresponding to a LSP with a strong higgsino
component. Such solutions are hardly realized in strongly constrained
models like mSUGRA.

Nevertheless, as every mSUGRA parameter set is contained in the full
MSSM, it is important to check that the additional MSSM parameters do
not have a large effect on the predictions for the observables and the
results of the minimization procedure. For example, the MSSM stau
co-annihilation point is similar to the corresponding mSUGRA point:
the gaugino mass parameters $\Mone$ and $\Mtwo$ are the values
obtained after the renormalization group evolution from the GUT scale
to the electroweak scale, as expected.  The MSSM generalization of the
$A$-funnel region shows a similar behavior.  The most sensitive
measurements are the Higgs boson mass and $\omegacdm$, and both are
within the theoretical error band. In the $h$-funnel scenario,
$\omegacdm$ is very sensitive to the exact value of the Higgs boson
mass, the change to the fixed MSSM parameters leads to a change of
150~MeV of the Higgs mass and an increase of $\omegacdm$.  The
bino--higgsino region does not exist for the mSUGRA model. It is a
generalization of the $A$-funnel, including chargino co-annihilation
via a charged Higgs in the $s$-channel. Essentially mass-degenerate
light neutralinos and charginos only appear for light winos or light
higgsinos, both not within the range of the renormalization group
equations starting from degenerate gaugino masses. For the same reason
the higgsino LSP point with chargino co-annihilation through gauge
bosons and into light quarks is also absent in the simplified mSUGRA
model.\bigskip

Exploring the MSSM for negative values of $\Mone$ leads to similar structures
in the ($\Mone,\Mtwo$) and ($\Mone,\mu$) planes. To be precise, while
for the ($\Mone,\Mtwo$) plane we observe a mirror symmetry with respect to the $\Mtwo$ axis,
for the ($\Mone,\mu$) plane we see a symmetry with respect to a simultaneous 
change of sign of both $\Mone$ and $\mu$. This observation is corroborated 
by the study of the parameter sets of Table~\ref{table:mMSSMparam}: if 
only the sign of $\Mone$ is changed, the solution becomes less probable. If additionally
the sign of $\mu$ is inverted, the mirror solution is as good as the original one. 
As the neutralino mixing matrix depends on $\mu$ and $\Mone$, a simultaneous change of sign
is equivalent to an unobservable global phase for the solutions considered here.\bigskip

In Figure~\ref{fig:MSSMm1m2m1muWmap} we show the same parameter
constraints as before, but for the WMAP measurement of the relic
density.  In the ($\Mone,\Mtwo$) plane only a hint of a difference is
visible, as WMAP allows for slightly lower $\Mtwo$.  In the
($\Mone,\mu$) plane, WMAP is more compatible in a slightly wider range
than Planck with the thin bino--higgsino region identified in
Figure~\ref{fig:MSSMm1m2m1muPlanck} (right). On the other hand, WMAP
gives slightly looser constraints for larger $\mu$ in the higgsino LSP
scenario for negative $\mu$. In addition, the $h$-funnel region is
less constrained by the WMAP measurement than by Planck.\bigskip

%--------------------------------------------------
\begin{figure}[t]
  \includegraphics[width=0.49\textwidth]{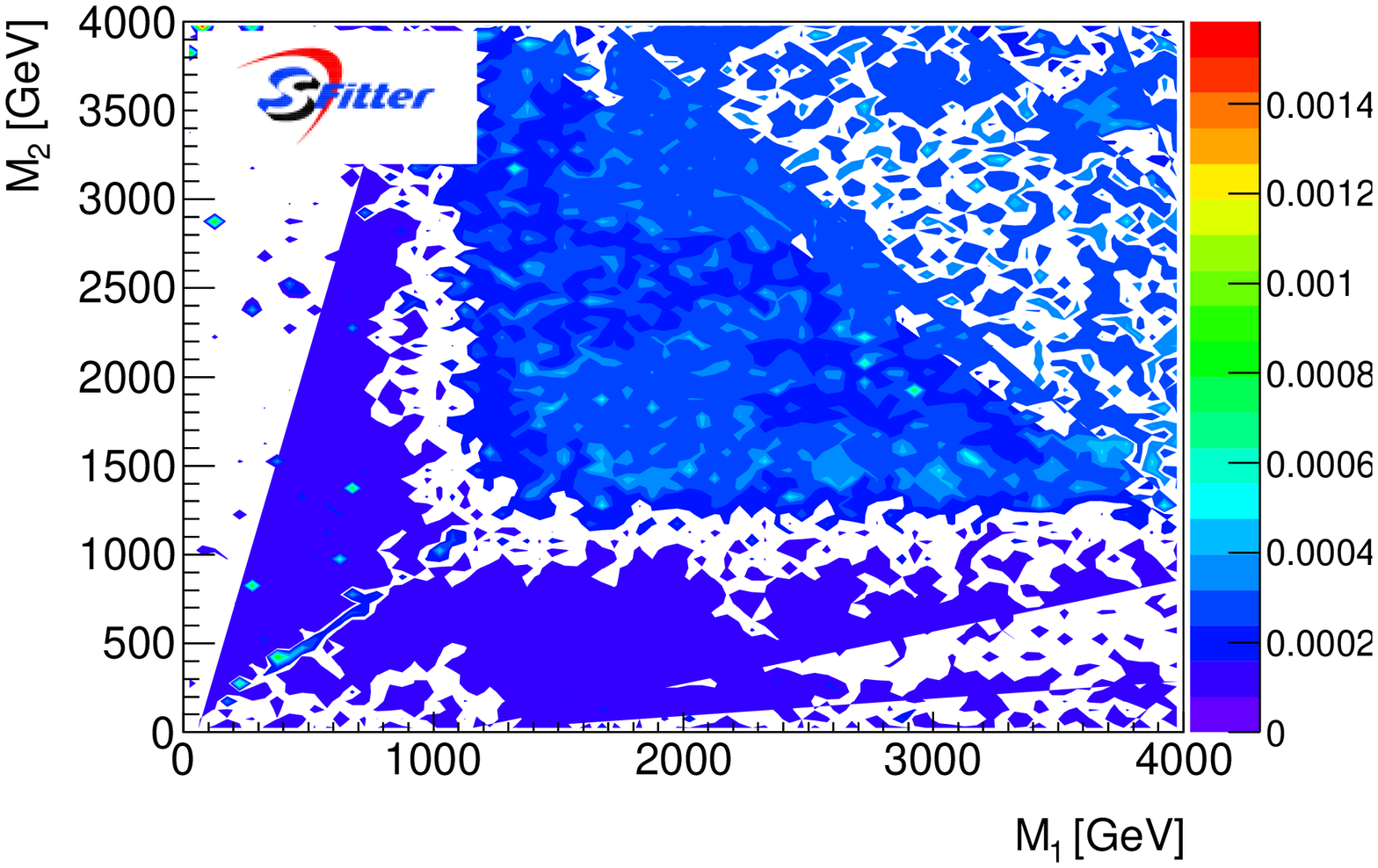}
  \includegraphics[width=0.49\textwidth]{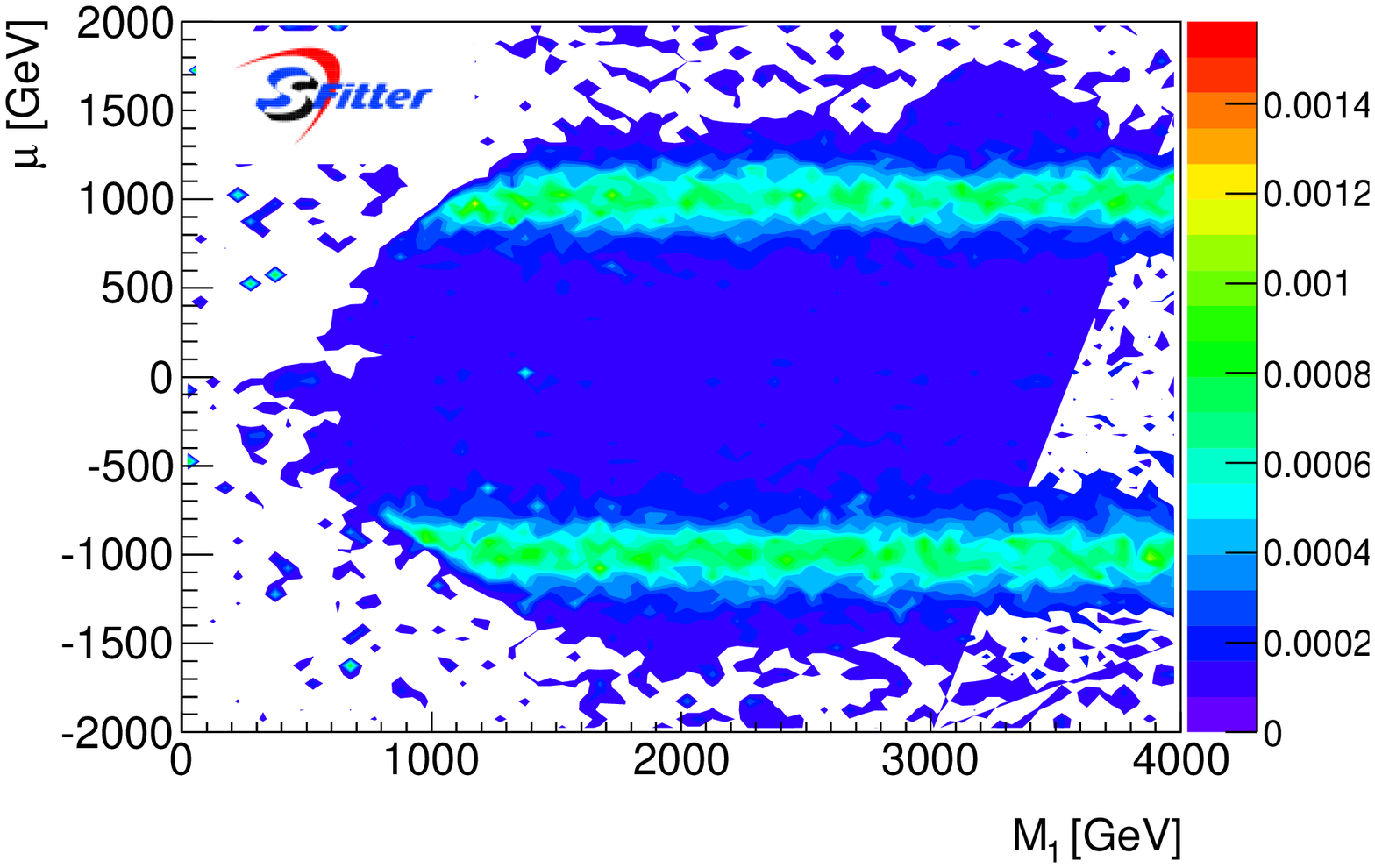}
  \caption{\label{fig:MSSMm1m2m1muBayes} Bayesian projections onto the
    ($\Mone,\Mtwo$) plane (left) and the ($\Mone,\mu$) plane using the
    Planck results combined with a $\tanb$--flat prior.}
\end{figure}
%--------------------------------------------------

As for the mSUGRA analysis, we also compare the profile likelihood
with a Bayesian approach. Volume effects can now affect the
determination of the model parameters, particularly changing the
balance between small and large parameter regions like the $h$-funnel
vs the higgsino regime.  As shown in
Figure~\ref{fig:MSSMm1m2m1muBayes}, the higgsino LSP region is indeed
identified the same way as in the Frequentist projection. The other
solutions are more sensitive to volume effects and therefore washed
out.  

%%%%%%%%%%%%%%%%%%%%%%%%%%%%%%%%%%%%%%%%%%%%%%%%%%%%%%%%%%%%%%%%%%%%%%%%%%%%%%%%
\section{Outlook}

Using \textsc{SFitter} we have studied the impact of measurements coming
from cosmological studies ($\omegacdm$), direct dark matter searches
(Xenon100), and collider measurements (Higgs mass) on the parameter
space of the mSUGRA model and on the TeV-scale MSSM. Additional direct
and indirect constraints have been included in the analysis, but turned out to
be secondary in defining the features of the preferred parameter
regions. 

We have compared the impact of the measurements of the dark matter relic density by Planck and
by WMAP, indicating a very slight shift in the best--fitting parameter
points. In contrast, a comparison of profile likelihood and Bayesian
methods to reduce the multi--dimensional parameter space showed
significant differences, arising from volume effects and choice of
prior. The latter can be chosen either at the GUT scale or at the TeV
scale, giving rise to a Jacobian scaling like $\tan^2 \beta$.\bigskip

The allowed regions of supersymmetric parameter space can best be
categorized by the dark matter annihilation channel. In mSUGRA we found
two valid regions, a narrow stau co-annihilation region at moderate
$\tanb$ and a large $A$-funnel region. Stop co-annihilation survives
the light Higgs mass constraint, but resides outside our tested range
of model parameter space, while the focus-point region seems to be
ruled out. 

In the TeV-scale MSSM we found narrow allowed regions corresponding to
stau co-annihilations and the light--Higgs funnel annihilation. The
heavy Higgs funnel becomes part of a large parameter region where the
lightest neutralino is a mixed bino--higgsino state, annihilating to
third--generation fermions. Chargino co-annihilation occurs with a
charged Higgs funnel. In addition, we observed a large higgsino region with
chargino and neutralino co-annihilation through gauge boson and into
light--flavor quarks. Finally, stop co-annihilation again resides
outside our range of model parameters. 

Because the allowed regions are very different in size, the Bayesian
analysis becomes sensitive to volume effects in comparing dark matter
annihilation channels. Moreover, in the light of these categories it
is not clear how we would define a simple effective theory covering
all these different supersymmetric scenarios, pointing towards a more
complex set of effective dark matter models.\bigskip

In terms of the supersymmetric Lagrangian we found that the positive
measurements like the relic density or the Higgs mass generally push
supersymmetry toward a high new physics mass scale. The absence of
signals for new physics at the 8~TeV run of the LHC puts little
tension into the parameter analysis. Nevertheless, several of the
parameter regions corresponding to different dark matter annihilation
can be probed by the LHC running at 13~TeV.  While there is a generic
benefit to testing a large variety of dark matter models, the successful
simple mSUGRA analysis indicates that there is no immediate need for
abandoning the standard WIMP hypothesis for the upcoming LHC
run.\bigskip

\begin{center}
{\bf Acknowledgments}
\end{center}

%\acknowledgments

We would like to thank \textsc{Fittino} for the stimulating
discussions we have had over the past years.  We are grateful to
Jean-Loic Kneur, Gilbert Moultaka and Michael Ughetto for discussions
on the supersymmetric spectrum calculation. 
We would also like to thank Laurent Duflot for his diligent reading 
and fruitful comments.
TP would like to thank the
CCPP at New York University for their hospitality while this paper was
finalized. He also wants to thank Neal Weiner for his continuous
encouragement to work on an mSUGRA analysis.
MR acknowledges partial support by the Deutsche Forschungsgemeinschaft
via the Sonderforschungsbereich/Transregio SFB/TR-9 ``Computational
Particle Physics'' and the Initiative and Networking Fund of the
Helmholtz Association, contract HA-101 (``Physics at the Terascale'').

%%%%%%%%%%%%%%%%%%%%%%%%%%%%%%%%%%%%%%%%%%%%%%%%%%%%%%%%%%%%%%%%%%%%%%%%%%%%%%%%

\end{document}